\documentclass[preprint]{elsarticle}
\usepackage[textwidth=18cm, textheight=24cm]{geometry}

\usepackage[utf8]{inputenc}
\usepackage[english]{babel}
\usepackage[T1]{fontenc}
\usepackage[hidelinks]{hyperref}
\usepackage{amsmath}
\usepackage{amssymb}
\usepackage{subcaption}
\usepackage{bm}
\usepackage{siunitx}
\usepackage{setspace}
\usepackage{cancel}
\usepackage{nicefrac}

\hypersetup{pdfauthor=author}

\def\etal.{et\penalty50\ al.}

\makeatletter
\def\ps@pprintTitle{%
  \let\@oddhead\@empty
  \let\@evenhead\@empty
  \let\@oddfoot\@empty
  \let\@evenfoot\@oddfoot
}
\makeatother

\begin{document}

\begin{frontmatter}
\title{Variational assimilation of sparse time-averaged data for efficient adjoint-based optimization of unsteady RANS simulations}


\author[ifd,empa]{Justin Plogmann\corref{corauthor}}
\cortext[corauthor]{Corresponding author}
\ead{justin.plogmann@empa.ch}

\author[ifd]{Oliver Brenner}
\ead{brennero@ethz.ch}

\author[ifd]{Patrick Jenny}
\ead{jenny@ifd.mavt.ethz.ch}

\affiliation[ifd]{%
    organization={Institute of Fluid Dynamics, ETH Zurich},%
    addressline={Sonneggstrasse~3},%
    city={Zürich},%
    postcode={CH-8092},%
    country={Switzerland}%
}

\affiliation[empa]{%
    organization={Chemical Energy Carriers and Vehicle Systems Laboratory, Swiss Federal Laboratories for Materials Science and Technology (Empa)},%
    addressline={Überlandstrasse~129},%
    city={Dübendorf},%
    postcode={CH-8600},%
    country={Switzerland}%
}

\begin{abstract}

Data assimilation (DA) plays a crucial role in extracting valuable information from flow measurements in fluid dynamics problems. Often only time-averaged data is available, which poses challenges for DA in the context of unsteady flow problems. Recent works have shown promising results in optimizing Reynolds-averaged Navier--Stokes (RANS) simulations of stationary flows using sparse data through variational data assimilation, enabling the reconstruction of mean flow profiles.

In this study we perform three-dimensional variational data assimilation of sparse time-averaged data into an unsteady RANS (URANS) simulation by means of a stationary divergence-free forcing term in the URANS equations. Efficiency and speed of our method are enhanced by employing coarse URANS simulations and leveraging the stationary discrete adjoint method for the time-averaged URANS equations. The data assimilation codes were developed in-house using OpenFOAM for the URANS simulations as well as for the solution of the adjoint problem, and Python for the gradient-based optimization.

Our results demonstrate that data assimilation of sparse time-averaged velocity measurements not only enables accurate mean flow reconstruction, but also improves the flow dynamics, specifically the vortex shedding frequency. To validate the efficacy of our approach, we applied it to turbulent flows around cylinders of various shapes at Reynolds numbers ranging from 3000 to 22000. Our findings indicate that data points near the cylinder play a crucial role in improving the vortex shedding frequency, while additional data points further downstream are necessary to also reconstruct the time-averaged velocity field in the wake region.

\end{abstract}

\begin{keyword}
Optimization \sep Data assimilation \sep CFD \sep URANS \sep Vortex shedding \sep Divergence-free \sep OpenFOAM 
\end{keyword}

\end{frontmatter}


\section{Introduction}
\label{sec:intro}
Simulations of complex flow problems remain a challenge, since they often occur at high Reynolds numbers. Therefore direct numerical simulations (DNS) and wall-resolved large eddy simulations (LES) are not feasible, since their computational cost scales with $Re^{3}$ and $Re^{1.8}$, respectively~\cite{Pope2000}.

In this study, the unsteady Reynolds-averaged Navier--Stokes (URANS) equations are employed. This approach models turbulent scales, while directly representing organized unsteady vortex formations. These vortex structures exhibit coherence due to their consistent occurrence. This property positions URANS as a technique capable of resolving these vortices, as discussed by Fröhlich and Terzi~\cite{Frohlich2008}. Consequently, when flow separates around a bluff body like a cylinder, it generates numerous smaller turbulent vortices along with recurring large ones. All turbulent fluctuations in the flow are addressed through the employed turbulence model. In contrast to URANS, the RANS approach lacks the capability to depict these time-varying coherent structures, as it lacks temporal resolution and is limited to describing statistically stationary conditions. Ultimately, both URANS and RANS outcomes represent expected values of the average flow, rooted in statistical analysis. Nevertheless, empirical evidence supports the superiority of URANS in accuracy and its ability to offer a more precise estimate of the mean flow, as highlighted in~\cite{Durbin1995}.

In any scenario, enhancing the predictability of current eddy viscosity (EV) models remains an imperative objective. In general, there exist two sources of errors, namely parametric and model-form errors~\cite{xiao19}. The parametric error emerges from the uncertainty surrounding the values of model coefficients~\cite{Pope2000}, but is not further considered here. 

Model-errors can be traced back to three distinct modeling aspects: the process of averaging, the functional depiction of the Reynolds stress term, and the selection of particular functions within the model itself~\cite{duraisamy19,XIAO17}. While errors tied to the initial two aspects are essentially unavoidable in EV models, enhancing the model's precision can be achieved through more astute choices of functions embedded within the model framework. 

Foures~\etal.~\cite{foures14} used the adjoint method for data assimilation (DA) in the context of RANS models, where the Reynolds stress term in the momentum equation is replaced by a momentum forcing that is completely driven by data. In conjunction with a gradient-based optimizer they found an optimal forcing that aims to minimize the discrepancy between the simulation and measurement data. More recently, Li~\etal.~\cite{li22} have extended this approach for cases with only limited numbers of wall pressure measurements. Moreover, Brenner~\etal.~\cite{brenner22} used a discrete adjoint-based DA technique to take into account sparsely distributed reference data and to correct the eddy viscosity for specific problems. An efficient discrete adjoint approach was proposed to compute the cost function gradients at a considerably low cost. To acknowledge the fact that this method only can optimize in regions where the mean shear-rate, i.e., the average rate-of-strain tensor is non-zero, Brenner~\etal.~\cite{brenner23} used a divergence-free forcing term in the RANS momentum equations instead. Similarly, Patel~\etal.~\cite{patel_turbulence_2023} also introduced a divergence-free forcing term into the RANS equations and compared the effectiveness of sparse velocity measurement data assimilation using a variational approach and physics-informed neural networks (PINN). PINN were originally introduced by Raissi~\etal.~\cite{raissi18} to solve physical problems with neural networks instead of using conventional numerical partial differential equation (PDE) solvers. Similarly, Physics-informed machine learning (PIML) (e.\,g.~\cite{karniadakis21,wang17}) aims to integrate physical knowledge into machine learning methods, such as neural networks or kernel-based methods.

Elsewhere, Fidkowski~\cite{FIDKOWSKI22} approximates the adjoint from a time-averaged primal by training machine learning models without full-state storage or checkpointing. Sliwinski and Rigas~\cite{sliwinski_mean_2023} performed PINN-based reconstruction of time-averaged quantities of an unsteady flow from sparse velocity data to infer the averaged unsteady forcing in the RANS equations.  Since the stationary RANS equations are enforced through the loss function in the PINN-based framework, no influence could be taken on the unsteady flow behavior. Only averaged unsteady second-order statistics, such as Reynolds stresses, were successfully recovered by inferring the averaged unsteady RANS forcing. Therefore, most DA studies concerning a dynamic (flow) behavior encounter the same problem. Either stationary (RANS) equations are used to reach a steady-state and time-averaged data can be assimilated to reconstruct and optimize the mean flow, or the traditional mathematical formulation in the time domain is exploited in conjunction with time-resolved observations. On one hand, as pointed out earlier, RANS simulations of wake flows provide no insight into the unsteadiness of the flow and, besides, often are not converging to a steady-state, but reach an unphysical and purely numerical shedding-state~\cite{schwarze_cfd-modellierung_2013}. On the other hand, however, two major drawbacks arise when using a dynamic DA approach. Firstly, time-resolved data needs to be available, but often only time-averaged data or first- and second-order statistics of the flow are provided. Secondly, it incurs huge computational cost, since time-stepping schemes within the DA procedure must be employed. Therefore, Koltukluoğlu~\cite{shen_fourier_2019} proposed a Fourier spectral dynamic DA framework that relies on the harmonically balanced momentum equations expressed in the frequency domain. The combination of the corresponding solutions yields the periodic solution of the original problem. Despite the success in reducing the wall clock times quite drastically, time-resolved data still is required.

In this paper we propose assimilating time-averaged data of unsteady flow problems into corresponding URANS simulations. To leverage the stationary variational DA technique at a low computational cost, time-averaging of the URANS equations is required. The DA technique incorporates time-averaged sparsely distributed reference velocity data obtained from a high fidelity simulation or measurement into the model to match the mean flow profiles as close as possible. The success of the DA method in matching time-averaged velocity data implies that the mean flow of unsteady flows can efficiently be reconstructed. In addition, if an unsteady flow characteristic, e.g. the Strouhal number, is optimized, we argue that the flow dynamics, specifically the vortex shedding frequency also gets improved. 

Therefore we utilize the discrete adjoint method for data assimilation, aiming at minimizing a cost function that quantifies the difference between the results of the forward model and the reference data. This method efficiently computes the gradient of the cost function with respect to the parameters. To find the cost function's minimum, we employ a gradient-based iterative optimization technique \cite{asch16}.

The primary reason for selecting the variational discrete adjoint method over other options is its ability to handle problems with large parameter sets. In our case, the number of parameters matches the number of mesh cells. Notably, certain non-intrusive black-box methods, like the ensemble Kalman filter, encounter the curse of dimensionality \cite{singh16}.

While the nudging or Newton relaxation method presents a computationally inexpensive data assimilation approach, it is limited to correcting the state variable field and not the parameter field. Moreover, if only a few reference points are used, it leads to a sub-optimal and noisy state variable field. On the other hand, the adjoint method, besides its application to data assimilation for closure problems, has found numerous other applications \cite{brenner22}.

The adjoint method can be implemented in continuous or discrete form. In the continuous approach, the governing equations are first differentiated, and then the resulting sensitivity equations are discretized before solving them numerically. Conversely, in the discrete approach, the governing equations are discretized first, and then differentiation yields sensitivity equations, which are subsequently solved numerically \cite{he18a}. We have opted for the discrete method due to its simplicity in constructing and implementing the discrete adjoint equations, as well as due to its natural handling of boundary conditions \cite{wang19}. Additionally, our codes were developed using the open-source computational fluid dynamics (CFD) toolbox \textit{foam-extend-5.0}, a version of OpenFOAM, where the discrete implementation is straightforward, thanks to the availability of various operators and tools acting on discrete equations that can be reused. However, discretization errors can affect the accuracy of the gradients, and the discrete adjoint method is sensitive to the choice of discretization schemes. Here, the continuous adjoint method tends to be less sensitive to the choice of mesh resolution and theoretically more accurate as it avoids the numerical errors associated with discretization.

\subsection{Objective and novelty of the present work}
\label{sec:Objective and novelty of the present work}

In this paper, we study the application of discrete adjoint-based data assimilation of time-averaged velocity reference data to correct URANS-based turbulent flow simulations with a stationary divergence-free forcing term. To utilize the stationary discrete adjoint method, temporal averaging of the URANS equations is employed. We demonstrate the method by applying it to flows around cylinders of various shapes at Reynolds numbers ranging from 3000 to 22000. In the computation of the cost function gradient in the discrete adjoint method, the evaluation of the derivative of the residual with respect to the parameters makes use of our efficient semi-analytical approach \cite{brenner22}. In addition, coarse URANS simulations make the optimization even more affordable. To the best of our knowledge, this work shows for the first time that the assimilation of sparse time-averaged velocity reference data into the URANS equations does not only allow for mean flow reconstruction, but also improves the dynamics of the unsteady flow solution, i.e., the vortex shedding frequency. The reference data distribution plays a crucial role and is further analyzed. The remainder of the paper is organized as follows. The method is introduced in Section \ref{sec:methods} and results for flows over cylinders of three different shapes at different Reynolds numbers are presented and discussed in Section \ref{sec:results}. Finally, in Section \ref{sec:conclusion}, the work is summarized, and future developments are suggested.


\section{Methods}
\label{sec:methods}

\subsection{Problem statement}
\label{sec: Problem statement}

In the following it is elaborated on the unsteady simulation of incompressible flow problems. Additionally, the stationary discrete adjoint method is applied to such flow problems to tune the stationary parameter in the context of data assimilation.

\subsubsection{Unsteady Reynolds-averaged Navier--Stokes equations}
\label{sec:Unsteady Reynolds-averaged Navier--Stokes equations}

To describe a turbulent flow with persistent unsteadiness, e.g. wake flows, URANS simulations are widely used in many fields of applications. The Reynolds decomposition of quantity $\xi$ is given by an
average $\bar{\xi}$ and a fluctuation $\xi'$. The unsteady RANS equation reads
\begin{equation}
    \label{eq:urans_momentum_no_assumption}
    \frac{\partial \bar{u}_{i}}{\partial t}
    +
    \frac{\partial \bar{u}_{i} \bar{u}_{j}}{\partial x_{j}}
    +
    \frac{\partial}{\partial x_{i}}
    \left[
        \frac{\bar{p}}{\rho}
    \right]
    -
    \frac{\partial}{\partial x_{j}}
    \left[
        2 \nu \bar{S}_{ij}
    \right]
    +
    \frac{\partial \overline{u'_{i} u'_{j}} }{\partial x_{j}}
    = 0
\end{equation}
with $\rho = \mathrm{const.}$ and the mean rate-of-strain tensor
\begin{equation}
    \bar{S}_{ij}
    =
    \frac{1}{2}
    \left(
        \frac{\partial \bar{u}_{i}}{\partial x_{j}}
        +
        \frac{\partial \bar{u}_{j}}{\partial x_{i}}
    \right)
    \ .
\end{equation}
The Reynolds stresses are modeled using the Boussinesq hypothesis, which introduces the eddy viscosity $\nu_t$.

\subsubsection{Data assimilation parameter}
\label{sec:Data assimilation parameter}

We add a correction term $f_{i}$ to account for discrepancies in the modeled Reynolds stresses as
\begin{equation}
    \frac{\partial \overline{u'_{i} u'_{j}}}{\partial x_{j}}
    =
    \frac{\partial}{\partial x_{j}}
    \left(
        \frac{2}{3}k\delta_{ij}
        -
        2\nu_{t}\bar{S}_{ij}
    \right)
    -
    f_{i}
    .
\end{equation}
This forcing $f_i$ is thus introduced on the right hand side of the momentum equation and then subjected to a Stokes--Helmholtz decomposition, similarly done in~\cite{foures14,li22,patel_turbulence_2023,sliwinski_mean_2023,perot_turbulence_1999,li_unsteady_2023}, i.\,e.,
\begin{equation}
    f_{i}
    =
    -\frac{\partial \phi}{\partial x_{i}}
    +
    \epsilon_{ijk} \frac{\partial \psi_{k}}{\partial x_{j}}
    ,
\end{equation}
with the scalar potential $\phi$, the vector potential $\psi$, and the Levi-Civita symbol $\epsilon_{ijk}$.

For the residual of the momentum equation this yields
\begin{equation}
    \label{eq:rans_momentum}
    R_{\bar{u}_{i}}
    =
    \frac{\partial \bar{u}_{i}}{\partial t}
    +
    \frac{\partial \bar{u}_{i} \bar{u}_{j}}{\partial x_{j}}
    +
    \frac{\partial p^{*}}{\partial x_{i}}
    -
    \frac{\partial}{\partial x_{j}}
    \left[
        2\nu_{\mathrm{eff}}\bar{S}_{ij}
    \right]
    -
    \epsilon_{ijk} \frac{\partial \psi_{k}}{\partial x_{j}}
    =
    0
    \ ,
\end{equation}
with effective viscosity
\begin{equation}
    \nu_{\mathrm{eff}}
    =
    \nu
    +
    \nu_{t}
    \ ,
\end{equation}
and where the averaged pressure $\bar{p}$, the isotropic part of the Reynolds stress tensor, and the scalar potential $\phi$ are absorbed into the modified pressure $p^{*}$ as
\begin{equation}
    \label{eq:rans_pressure_mod}
    p^{*}
    =
    \frac{\bar{p}}{\rho}
    +
    \frac{2}{3} k
    +
    \phi
    \ .
\end{equation}

To simplify the notation in the remainder of the manuscript, $\bar{p}$ is used for the modified pressure $p^{*}$. Therefore, only the vector potential $\psi_{k}$ appears explicitly in the momentum equations. Further, data assimilation is directly acting on $\psi_{k}$, such that no additional equations need to be solved for $\phi$ and $\psi_{k}$. For two-dimensional simulations only the $\psi_{z}$ component is relevant, i.\,e., a scalar field is sufficient to influence both velocity components, which reduces complexity over directly tuning a force $f_{i}$. Since $\psi_{z}$ is a non-uniform field linked to the dimensions of the computational mesh, it results in $N$ degrees of freedom in the parameter space for two-dimensional flows, with $N$ being the number of mesh cells. For more details it is referred to \cite{brenner23}. 

\subsubsection{Temporal averaging of the URANS equations}
\label{sec:Temporal averaging of URANS equations}

The goal for this work is to leverage the discrete adjoint method for stationary flows from \cite{brenner23} and apply it to unsteady flow problems. Therefore we introduce temporal averaging $\left<\cdot\right>$ with the corresponding fluctuation $\left(\cdot\right)''$ as
\begin{equation}
    \bar{\xi}
    = \left<\bar{\xi}\right>
    + \bar{\xi}''
    ,
\end{equation}
when applied to a quantity $\bar{\xi}$ that already is Reynolds- or ensemble-averaged. A detailed derivation of time-averaging the URANS equations is given in \ref{sec:Temporal averaging of URANS equations appendix}. With all terms expanded and reordered, we obtain for the momentum equation
\begin{equation}
\label{eq:urans_time_average}
    R_{\left<\bar{u}_{i}\right>}
    =
    \frac{\partial \left<\bar{u}_{i}\right> \left<\bar{u}_{j}\right>}{\partial x_{j}}
     + \frac{\partial \left<\bar{p}\right>}{\partial x_{i}}
    - \frac{\partial}{\partial x_{j}}
        \left(
            2 \left(\nu + \left<\nu_{t}\right>\right) \left<\bar{S}_{ij} \right>
        \right) 
        - \underbrace{\frac{\partial}{\partial x_{j}}
        \left(
            2 \left<\nu''_{t} \bar{S}_{ij}'' \right>
        \right)
     + \frac{\partial \left<\bar{u}''_{i} \bar{u}''_{j}\right>}{\partial x_{j}}}_{\mathrm{additional~stress~terms}}
     - \epsilon_{ijk} \frac{\partial \psi_{k}}{\partial x_{j}}
     = 0.
\end{equation}

This equation is very similar to the original URANS equations except that it describes a stationary state and that two additional terms appeared due to the time-averaging process, which are tracked during the forward problem solution. The discrete adjoint method now can be used the same way it is used in \cite{brenner23} for RANS, but for the time-averaged URANS equations \eqref{eq:urans_time_average}. To ensure converged averaged properties of the flow fields that are needed in conjunction with their corresponding fluctuations to construct Eq.~\eqref{eq:urans_time_average}, we introduce a global measure for the change in an averaged field, in this case the velocity, between time steps $(i-1)$ and $(i)$ as
\begin{equation}
\label{eq:averaging convergence criterion}
    C_U = \int_\Omega \frac{\delta \lVert \langle \bar{u} \rangle \rVert}{\delta t} \mathrm{d}V = \int_\Omega \frac{\lVert \langle \bar{u} \rangle ^{(i)} - \langle \bar{u} \rangle ^{(i-1)} \rVert}{\Delta t ^{(i)}} \mathrm{d}V \approx \sum_k \frac{\lVert \langle \bar{u} \rangle _k ^{(i)} - \langle \bar{u} \rangle _k ^{(i-1)} \rVert}{\Delta t ^{(i)}} V_k 
\end{equation}
with the velocity vector $\langle \bar{u}\rangle_k$, the vector norm $\lVert \cdot \rVert$, the time step size $\Delta t$ and the cell volume size $V_k$. The criterion is set such that the averaged quantities are well converged, but the number of time steps needed in the forward problem solution is kept as low as possible to minimize the required CPU time. Therefore the criterion is case-dependent and set individually for each case as discussed in Section \ref{sec:results}. 

\subsection{Spectral analysis}
\label{sec:Spectral analysis}

To validate the effectiveness of our data assimilation approach for unsteady flows, we perform a spectral analysis to study the dynamic flow behavior before and after the optimization and compare it with the corresponding reference. For this purpose the lift coefficient
\begin{equation}
    \label{eq:lift coefficient}
    c_l = \frac{2F_l}{\rho u_\infty^2 A} 
\end{equation}
with lift force $F_l$, density $\rho$, free-stream velocity $u_\infty$ and projected area $A$ is computed. The time signal is sampled with a uniform time-step during the forward problem (URANS) computation for a sufficient number of periods, which is 50 here. Additionally, the initial transient, where the solution of the forward problem is not yet converged is omitted in the time signal for the spectral analysis. The frequency obtained from this analysis is considered the Strouhal frequency $f_s$ or first harmonic. Often this frequency governs the dominant vortex shedding in the wake. URANS simulations are not able to resolve any turbulent structures but can represent the coherent vortex structures associated with the Strouhal frequency. Therefore, higher harmonics are not investigated throughout this work as well as the restoration of spectral features are not considered as part of this research. Sometimes, however, the second harmonic is associated with the dominant vortex shedding in the wake \cite{trias_turbulent_2015}. This depends on the flow conditions, cylinder geometry and spatial location in the wake. Nevertheless, in this work the Strouhal number 
\begin{equation}
    St = \frac{f_s D}{u_\infty}
\end{equation}
with the cylinder dimension $D$ is used as a global measure for the flow dynamics and is compared to the respective reference Strouhal number.

\subsection{Data assimilation problem}
\label{sec:Data assimilation problem}

The data assimilation problem requires to minimize the discrepancy between the state variables computed by the RANS model and the existing reference data and thus can be constructed as an optimization problem. In this work the scalar cost function $f$ consists of a regularization function $f_{\psi}$ and a discrepancy contribution $f_{U}$, i.\,e.,
\begin{equation}
    \label{eq:cost_function}
    f\left(\psi, U\right)
    =
    f_{\psi}\left(\psi\right)
    +
    f_{U}\left(U\right)
    .
\end{equation}
It is subject to the residual $R$ of the governing equation, that is, one seeks
\begin{subequations}
    \begin{alignat}{2}
        \label{eq:minimization_problem}
        & \!\min_{\psi}     & \quad & f\left(\psi, U\right) \\
        \label{eq:minimization_problem_constr_1}
        & \text{subject to} &       & R\left(\psi, U\right) = 0 \ ,
    \end{alignat}
\end{subequations}
where $\psi$ is the parameter to be optimized and $U$ the forward problem solution (not limited to the velocity).

The optimization involves an inverse problem, which is highly non-linear and usually underdetermined. Hence, a non-linear optimization solver is used, but no assurance is given that there exists a unique solution~\cite{foures14}. Therefore, some form of regularization is introduced to reduce the ambiguity. In addition, the URANS equations serve as a physical model providing a good initial solution such that the discrepancy between the initial and reference solution is already small. Hence, only minor corrections have to be made in order to reach the minimum. For each data assimilation iteration a parameter update is computed with a gradient descent approach. In the present work the optimization algorithm \emph{Demon Adam}~\cite{chen19} is chosen.

\subsubsection{Discrete adjoint method}
\label{sec:Discrete adjoint method}

In this paper, the stationary discrete adjoint technique is employed to calculate the gradient of the cost function concerning the parameters in question. This strategy demonstrates computational efficiency by yielding a cost of computing the gradient that is akin to solving the stationary forward system. Notably, the computational cost remains unaffected by the quantity of parameters involved. This characteristic becomes especially pertinent in cases like this, where the parameter field $\psi$ could encompass numerous degrees of freedom linked to the mesh resolution. However, a notable downside lies in the intricate nature of establishing the gradient computation, given its intrusive methodology. A summary including a derivation is provided in \cite{brenner22}. The latest modifications to the implementation and the transfer of the concept from a modified eddy viscosity approach used in \cite{brenner22} to the divergence-free forcing approach are described in \cite{brenner23}.

In the discrete case the forward problem is formulated as a system of linear equations $R$ with solution vector $U$ as defined in Eq.~\eqref{eq:solution vector}. This yields 
\begin{equation}
    \label{eq:coupled_adjoint}
    \left( \frac{\partial R}{\partial U} \right)^{T}
    \lambda
    =
    \left( \frac{\partial f_{U}}{\partial U} \right)^{T}
\end{equation}
as an equation for the Lagrange multipliers $\lambda$. Solving this additional equation and substituting the results in Eq.~\eqref{eq:adjoint_gradient} gives
\begin{equation}
    \label{eq:rans_adjoint_gradient}
    \begin{split}
         \frac{\mathrm{d} f}{\mathrm{d} \psi}
    &=
    \frac{\partial f}{\partial \psi}
    -
    \lambda^{T} \frac{\partial R}{\partial \psi} \\
    &=
    \frac{\partial f}{\partial \psi}
    -
    \lambda_{\left<\bar{u}_{x}\right>}^{T} \frac{\partial R_{\left<\bar{u}_{x}\right>}}{\partial \psi}
    -
    \lambda_{\left<\bar{u}_{y}\right>}^{T} \frac{\partial R_{\left<\bar{u}_{y}\right>}}{\partial \psi}
    -
    \lambda_{\left<\bar{u}_{z}\right>}^{T} \frac{\partial R_{\left<\bar{u}_{z}\right>}}{\partial \psi}
    -
    \lambda_{\left<\bar{p}\right>}^{T} \frac{\partial R_{\left<\bar{p}\right>}}{\partial \psi}
    \end{split}
\end{equation}
for the adjoint gradient. We suggest a method that relies on the discretization operators offered in this specific instance by OpenFOAM. The fundamental concept involves discretizing and linearizing the residual of the forward problem, denoted as $R$, concerning the variable under consideration, which either is the parameter field $\psi$ or the solution vector $U$.

For more details, please refer to \cite{brenner22} and \ref{sec:Discrete adjoint method appendix}.

\subsubsection{Cost function and regularization}
\label{sec:Cost function and regularization}

As stated in Eq.~\eqref{eq:cost_function}, the cost function consists of a regularization part and a discrepancy part. Here, total variation (TV) regularization is chosen to reduce ambiguity of the inverse problem. It is based on a measure for smoothness of the parameter field. Other works have focused on improving the regularization of inverse problems, since they always are underdetermined and thus ambiguous. Piroozmand~\etal.~\cite{piroozmand_dimensionality_2023} introduced a piecewise linear dimensionality reduction regularization method to constrain the corrective parameter by a piecewise linear variation to reduce ambiguity and avoid unphysical spikes in the solution field. Epp~\etal.~\cite{epp22} tackle the ambiguity problem by using a hierarchical regularization and employing physically meaningful constraints and thus avoiding further parameter tuning. However, the optimal regularization method is not the focus of this work, so TV regularization as proposed in~\cite{brenner22} is considered sufficient. Furthermore, the optimal placement of reference data points is not the focus of this work. In~\cite{brenner23}, the authors demonstrate in case of RANS simulations that reducing the number of reference data points does only marginally degrade the accuracy of mean flow reconstruction. This is only possible with appropriate regularization. If only a very few number of reference data points would be available, but no regularization is applied, the solution field might not be smooth but consists of unphysical spikes at the locations of these very few data points. Thus, accurate mean flow reconstruction would not be achieved in that case. However, if the regularization weight is set high enough, the solution will not be worse than the URANS baseline result since TV regularization will smooth out these spikes. Hence, one should be aware that the regularization weight is a crucial parameter in the optimization framework. 

The discrepancy part of the cost function measures agreement of the forward problem solution $U$ with the reference data $U^{\mathrm{ref}}$.
In the presented application only time-averaged velocity data is assimilated, i.\,e.,
\begin{equation}
    \label{eq:discrepancy}
    f_{U}\left(U\right)
    =
    \frac{
        1
    }{
        V^{\mathrm{ref}}
    }
    \sum\limits_{j\in\mathcal{R}} \left[
        \sum\limits_{k\in\left\{x,y,z\right\}}
        \left(
            \langle\bar{u}_{k,j}\rangle
            -
             \langle\bar{u}_{k,j}^{\mathrm{ref}}\rangle
        \right)^{2}
        V_{j}
    \right]
    \ ,
\end{equation}
where $\mathcal{R}$ is the list of reference cell indices $j$, $V_{j}$ the volume of cell $j$, and
\begin{equation}
    V^{\mathrm{ref}}
    =
    \sum\limits_{j\in\mathcal{R}}{V_{j}}
\end{equation}
the volume of all reference cells. Since the objective of this work is the mean flow reconstruction and assuming that the reference data error/noise is negligible, an uncertainty quantification has not been performed. Therefore, observation errors are neglected in the cost function. 

For two-dimensional flows, TV regularization is only applied to the $\psi_{z}$ component of the parameter field.
This is done without loss of generality, since the regularization can be applied to the other components as well.
In particular, following the method by Brenner~\etal.~\cite{brenner22}, a function of the form
\begin{equation}
    \label{eq:regularization}
    f_{\psi}\left(\psi\right)
    =
    C^{\mathrm{reg}}
    \sum\limits_{i\in\Omega} \left[
        \frac{1}{\left|\mathcal{B}_{i}\right|}
        \sum\limits_{k\in\mathcal{B}_{i}} \left(
            \psi_{z,i}
            -
            \psi_{z,k}
        \right)^{2}
    \right]
    \ ,
\end{equation}
with weight parameter $C^{\mathrm{reg}}$, is used to punish non-smooth parameter fields.
Here, index $i$ loops over all cells in the simulation domain $\Omega$ and index $k$ loops over $\mathcal{B}_{i}$ the list of indices of neighboring cells of cell $i$, with $\left|\mathcal{B}_{i}\right|$ denoting the number of neighboring cells of cell $i$.

\subsubsection{Test function}
\label{sec:test function}

We introduce another functional called test function $f^\mathrm{test}$ that is defined analogously to the discrepancy cost function. However, instead of evaluating the test function at the reference data points, test data points are introduced, where then $f^\mathrm{test}$ is calculated according to Eq.~\eqref{eq:discrepancy}. The test data points are all remaining points in the domain which are not defined as reference points. This concept often is considered in machine learning, where training data is used to fit the parameters of the model. After a validation data set, test data is used to provide an unbiased evaluation of a final model fit on the training data set. In our case, the test function illustrates, if the optimization based on the reference data points also is working in the remaining part of the domain. In addition, the test function is very important to tune the TV regularization, since overfitting or sharp changes would be penalized with a large contribution to the test function. Hence, the test function is an essential measure to make sure that the resulting fields after the optimization exhibit a smooth and noiseless spatial distribution.

\subsection{Implementation}
\label{sec:Implementation}

The open-source field operation and manipulation platform, OpenFOAM, often used for computational fluid dynamics (CFD) specifically the version \textit{foam-extend-5.0}~\cite{fe50}, is where the solvers are implemented, leveraging the platform's pre-existing solvers and diverse capabilities. The approach adopted for solving both the forward and adjoint problems involves a fully coupled solution process. Notably, the \texttt{transientFoam} solver was extended.

The computational meshes were created using \texttt{gmsh}~\cite{geuzaine09} and \texttt{snappyHexMesh}, and the cell size is decreasing toward the cylinder wall to capture the flow detachment well enough. However, the wall-normal distance still is quite large to conform with the requirements of wall functions. This makes the mesh relatively coarse and allows for larger time steps, which in turn decreases the CPU time of the solution of the forward problem drastically. This methodology of mesh generation is applied to all cases in this work.

Second order schemes are used for discretization, and a bi-conjugate gradient stabilized linear solver with a Cholesky preconditioner is used for the forward coupled system, whereas no preconditioning is applied to the adjoint coupled system. The implicit Euler scheme is chosen for time integration, while a constant time step is used for uniform sampling as part of the spectral analysis. The time step sizes for all cases are listed in Section~\ref{sec:results} ensuring an average Courant-Friedrichs-Lewy (CFL) number smaller than 1. For this particular work the two-equation $k$-$\omega$ SST model as proposed by \cite{Menter2003} is used due to its ability to account for the transport of the principal shear stress in adverse pressure gradient boundary layers~\cite{menterReviewShearstressTransport2009}. Moreover, OpenFOAM incorporates parallel computing as an inherent capability, utilizing domain decomposition. This involves dividing the geometry and its corresponding fields into segments, which are then assigned to individual processors. The parallel calculations are executed using the widely adopted \textit{openMP} implementation from the standard message passing interface (MPI) domain. Parallel computing is only used for the solution of the forward problem, but not for the adjoint problem solution.

The optimization itself is performed in Python having an interface to OpenFOAM using \texttt{pyFOAM}. Moreover, the reference data, which is averaged temporally and in $z$ direction is interpolated on the corresponding mesh using \texttt{griddata} from \texttt{SciPy}.

\begin{figure}[!ht]
    \centering
    \includegraphics[]{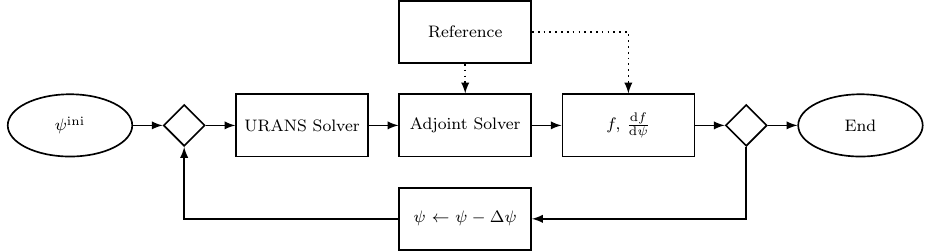}
    \caption{
        Flowchart of the optimization based data assimilation procedure.
        Starting from an initial URANS setup, the parameter field $\psi$ is updated iteratively based on the cost function gradient.
        A \textit{foam-extend-5.0} solver is used to evaluate the forward and adjoint problems, as well as the cost function and the gradient.
    }
    \label{fig:diagram_rans_da}
\end{figure}

The overall computational procedure as illustrated in Fig.~\ref{fig:diagram_rans_da} thus works as follows. First, the forward problem (URANS) is solved to obtain the forward solution $U$ as a function of the parameter field $\psi$. Meanwhile, time-averaging the URANS equations is introduced and the solver stops as soon as the averages converge according to the criterion for $C_U$. Then, the time-averaged and converged forward system matrix $\frac{\partial R}{\partial U}$ is used in conjunction with the analytical evaluation of the right-hand side of Eq.~\eqref{eq:coupled_adjoint} to solve the adjoint system for the Lagrange multipliers $\lambda$. Third, the matrix $\frac{\partial R}{\partial \psi}$ is constructed and, together with the regularization function and the Lagrange multipliers $\lambda$, used to compute the adjoint gradient. The current parameter values and the corresponding gradient are then used in the optimization step to update the parameter field.


\section{Results and discussion}
\label{sec:results}

In the following, the above described data assimilation framework is used to optimize flows around circular, square and rectangular cylinders. For this purpose, synthetic and DNS reference data from the literature is considered. 

\subsection{Validation of data assimilation procedure using synthetic reference data}
\label{sec:Validation with synthetic reference data}

\subsubsection{Test case setup of flow around two-dimensional circular cylinder}
\label{sec:Test case setup of flow around two-dimensional circular cylinder}

\begin{figure}[!ht]
    \centering
    \includegraphics[]{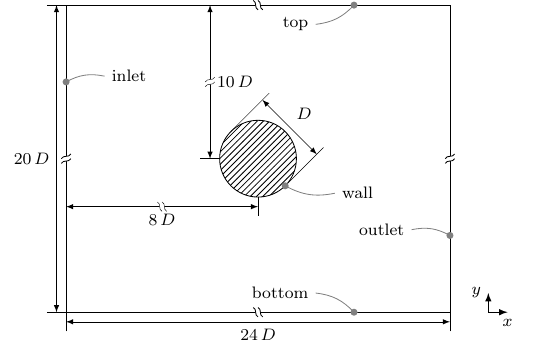}
    \caption{
            Simulation domain of the two-dimensional circular cylinder setup. Mean flow is in $x$ direction.
            All length scales are normalized by cylinder diameter $D$.
    }
    \label{fig:circular_cylinder_2d}
\end{figure}

The performance of our data assimilation approach is demonstrated for flow around a two-dimensional circular cylinder (e.g.~\cite{lehmkuhl_low-frequency_2013}). A sketch of the geometry and boundary conditions is provided in Fig.~\ref{fig:circular_cylinder_2d}. Depending on the Reynolds number, flow detachments occur and vortex shedding prevails in the wake of the cylinder. All length scales are expressed relative to the cylinder diameter $D$ and the Reynolds number is computed from $D$, the free-stream velocity $u_\infty$ and the kinematic viscosity $\nu$. Boundary conditions (BC's) for velocity and pressure are taken from \cite{lehmkuhl_low-frequency_2013}. Wall functions are applied at the cylinder wall, 2D BC's are present in spanwise directions and slip conditions are used for the top and bottom boundaries. At the inlet Dirichlet BC's are used and set to a small value reproducing the inflow conditions from \cite{lehmkuhl_low-frequency_2013}. Neumann BC's are set at the outlet. In this work we analyze the case for
\begin{equation*}
    \mathrm{Re}
    =
    \frac{u_{\infty}D}{\nu}
    =
    \num{3900}
\end{equation*}
with synthetic reference data generated in different ways (see Sections~\ref{sec:Synthetic reference data created with imposed divergence-free force} and \ref{sec:Synthetic reference data created by manipulating eddy viscosity }). A time step size of $\frac{\Delta t u_\infty}{D} = 0.1$ is used.

The computational mesh was created using \texttt{gmsh} and consists of 5700 hexahedral cells. The ratio of the smallest to largest cell is $7.614 \cdot 10^{-3}$.

\subsubsection{Synthetic reference data generated with imposed divergence-free force}
\label{sec:Synthetic reference data created with imposed divergence-free force}

\begin{figure}[!ht]
     \centering
     \begin{subfigure}[t]{0.49\textwidth}
         \centering
         \includegraphics[width=\textwidth]{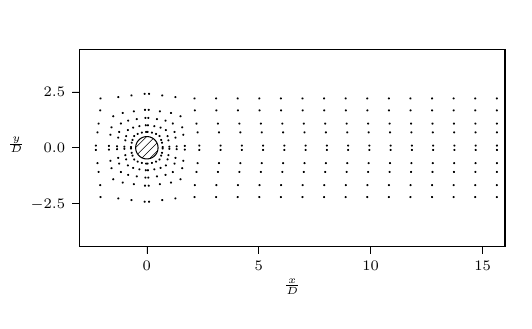}
         \caption{Reference data point locations around circular cylinder and its wake. In total 284 points are selected.}
         \label{fig:reference data points circular cylinder}
     \end{subfigure}
     \hfill
     \begin{subfigure}[t]{0.49\textwidth}
         \centering
         \includegraphics[width=\textwidth]{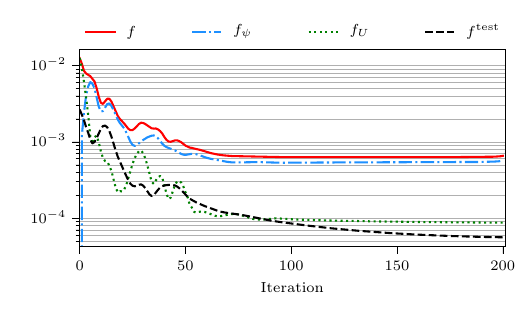}
         \caption{Cost function.}
         \label{fig:cost function circular cylinder curl psi ref}
     \end{subfigure}
     \hfill
     \begin{subfigure}[b]{0.49\textwidth}
         \centering
         \includegraphics[width=\textwidth]{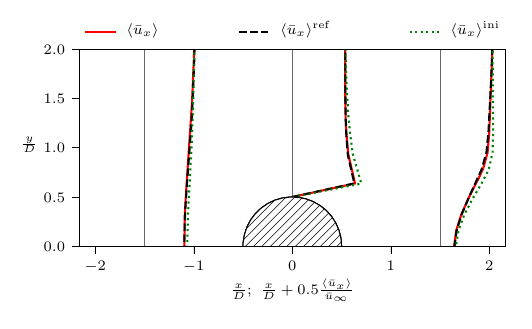}
         \caption{Mean flow velocity profiles near the circular cylinder.}
         \label{fig:velocity profiles circular cylinder psi}
     \end{subfigure}
     \hfill
     \begin{subfigure}[b]{0.49\textwidth}
         \centering
         \includegraphics[width=\textwidth]{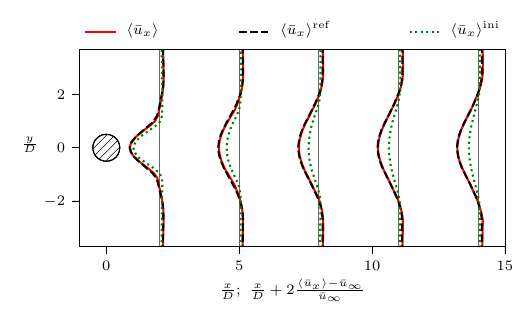}
         \caption{Mean flow velocity profiles in the wake of the circular cylinder.}
         \label{fig:velocity profiles circular cylinder wake psi}
     \end{subfigure}
        \caption{Optimization of flow around the two-dimensional circular cylinder using synthetic reference data created with an imposed divergence-free force. Optimization step size is $\eta=\num{5e-5}$ with a maximum number of optimization steps $N=200$. Convergence of the forward problem solver is reached after $C_U=0.1$. The regularization weight parameter is set to $C^{\mathrm{reg}}=\num{2e-1}$.}
        \label{fig:circular cylinder curl psi results}
\end{figure}

First, the default forward solver is used to solve the basic forward problem and to obtain the initial unperturbed (baseline) solution. Second, an arbitrary divergence-free forcing field $\nabla \times \psi_\mathrm{ref}$ is prescribed and used by the modified forward solver in conjunction with the initial solution. The resulting synthetic reference velocity field $\langle \bar{u} \rangle^\mathrm{ref}$ is then used as a reference for data assimilation. Third, the data assimilation code is run for the same setup, starting with $\nabla~\times~\psi^\mathrm{ref}=(0,0,0)$ everywhere. All cells are supplied with $\langle \bar{u}_x \rangle$ and $\langle \bar{u}_y \rangle$ reference data, and regularization is applied. The single objective is to test whether the synthetic velocity field can be recovered through the proposed data assimilation approach. The reference forcing field is chosen to be a smooth function for the $x$ and $y$ components as described by Eq.~\eqref{eq:curl psi synth ref} and also illustrated in Fig.~\ref{fig:synthetic ref curl psi creation}. As for the reference data points, the distribution depicted in Fig.~\ref{fig:reference data points circular cylinder} is chosen. Therefore, data points in the wake and proximity of the cylinder are present. This sparsity map is just one example of how to allocate data points. In general, Brenner~\etal.~\cite{brenner22} have shown that the optimization works for different densities of reference data point distributions. In addition, Brenner~\etal.~\cite{brenner23} have found a great robustness of the TV regularization parameters to the reference data point distribution densities.
As can be seen in Fig.~\ref{fig:cost function circular cylinder curl psi ref}, the cost function decreases drastically after a few optimization steps. Moreover, the test function decreases as a result of the regularization. Hence, the discrepancy between the optimized and synthetically generated velocity fields also diminishes in the remaining cells where no reference is given. The velocity profiles in the wake of the cylinder (see Fig.~\ref{fig:velocity profiles circular cylinder wake psi}) as well as near the cylinder (see Fig.~\ref{fig:velocity profiles circular cylinder psi}) illustrate the smoothness of the recovered velocity field and the improvement in terms of matching the synthetic reference.

As a result of the imposed forcing used to generate the synthetic reference data, a different vortex shedding frequency and hence Strouhal number was obtained. The baseline solution using the default forward problem solver yielded a Strouhal number of $St=0.195$. The synthetic reference velocity field, however, relates to $St=0.175$. After the assimilation of time-averaged synthetic velocity data, the vortex shedding frequency also changed, leading to $St=0.180$. This is a significant improvement considering that only the time-averaged data was assimilated. The stationary divergence-free forcing term obtained from the temporally averaged URANS equations~\eqref{eq:urans_time_average} acts again in the URANS equations~\eqref{eq:rans_momentum} of the forward problem solver and hence improves the flow dynamics. Nevertheless, the synthetic reference velocity field was created in the same way the optimization algorithm tries to recover it, i.e., with a divergence-free forcing term. Therefore, a different synthetic reference data generation procedure is employed in Section \ref{sec:Synthetic reference data created by manipulating eddy viscosity }.

\subsubsection{Synthetic reference data generated by local scaling of eddy viscosity}
\label{sec:Synthetic reference data created by manipulating eddy viscosity }

Another way to create synthetic reference data is to influence the turbulence equations, that is, manipulating the eddy viscosity by multiplying it with a scalar field $\alpha^\mathrm{ref}$. Hence, the eddy viscosity still is computed with the unmodified $k$-$\omega$ SST model and the scalar field $\alpha^\mathrm{ref}$ is subsequently used as a local scaling factor. The modified eddy viscosity differs from the originally computed eddy viscosity and thus has an effect on the velocity field. We introduce a smooth scalar field (see Eq.~\eqref{eq:alpha nut ref}) as depicted in Fig. \ref{fig:synthetic ref alpha ref creation}. 

\begin{figure}[!ht]
     \centering
     \begin{subfigure}[b]{\textwidth}
         \centering
         \includegraphics[width=0.49\textwidth]{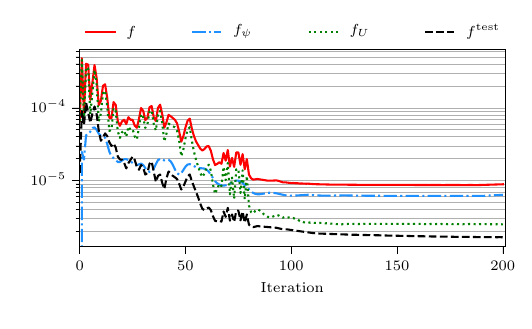}
         \caption{Cost function.}
         \label{fig:cost function circular cylinder alpha}
     \end{subfigure}
     \newline
     \begin{subfigure}[b]{0.49\textwidth}
         \centering
         \includegraphics[width=\textwidth]{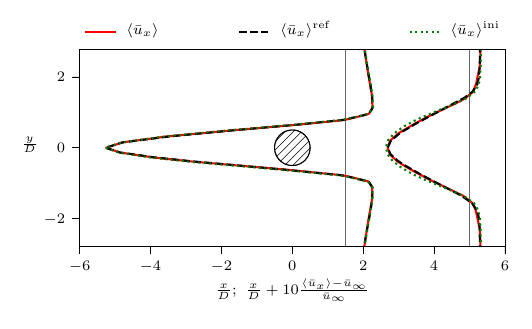}
         \caption{Mean flow velocity profiles in the near-wake region.}
         \label{fig:velocity profiles circular cylinder wake alpha}
     \end{subfigure}
     \hfill
     \begin{subfigure}[b]{0.49\textwidth}
         \centering
         \includegraphics[width=\textwidth]{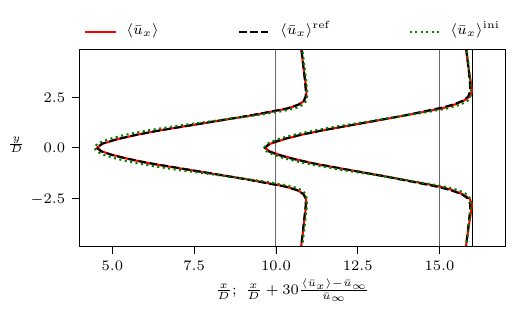}
         \caption{Mean flow velocity profiles in the far-wake region.}
         \label{fig:velocity profiles circular cylinder far wake alpha}
     \end{subfigure}
        \caption{Optimization of flow around the two-dimensional circular cylinder using synthetic reference data created by manipulating the eddy viscosity with a local scaling factor $\alpha^\mathrm{ref}$. Optimization step size is $\eta=\num{3e-5}$ with a maximum number of optimization steps $N=200$. Convergence of the forward problem solver is reached after $C_U=0.1$. The regularization weight parameter is set to $C^{\mathrm{reg}}=\num{1e-2}$.}
        \label{fig:circular cylinder alpha results}
\end{figure}

As Fig.~\ref{fig:cost function circular cylinder alpha} illustrates, the cost function mainly decreases within the first 100 iterations and then reaches a minimum. Regularization again was needed to also bring down the test function, that is, optimizing the velocity field in the remaining points of the domain. Figures~\ref{fig:velocity profiles circular cylinder wake alpha} and \ref{fig:velocity profiles circular cylinder far wake alpha} show that the velocity profiles of synthetic reference and initial solution do not differ that much. In fact, this is due to the design of the scalar field $\alpha^\mathrm{ref}$. A more pronounced field would have resulted in a larger deviation from the initial state. However, the results obtained with a higher magnitude of $\alpha^\mathrm{ref}$ or a larger spatial range were not physically meaningful. In particular, the velocity profiles showed unphysical shapes in the wake making them insufficient as reference data.  Thus, the scalar field described in Eq.~\eqref{eq:alpha nut ref} was chosen. Moreover, it demonstrates that the optimization algorithm can not only be used to minimize large discrepancies, but also works well to fine tune results. 

Consequently, the modified eddy viscosity used to generate the synthetic reference data resulted in a different Strouhal number compared to the Strouhal number of the baseline solution. The synthetic reference corresponds to $St=0.202$. After the optimization, the Strouhal number changed from $St=0.195$ to $St=0.200$ and is therefore almost perfectly coinciding with the reference Strouhal number.

In summary, the optimization works very well for both cases where synthetic reference data was created with the forward problem solver. The optimized velocity profiles show a very good match with the reference. This also results in a good agreement of the respective Strouhal numbers. Additionally, only a few reference points were selected and the results could be improved by a more optimal placement of the reference points. However, we demonstrated the effectiveness of assimilating time-averaged synthetic velocity data into URANS-based simulations by employing the stationary discrete adjoint method on the time-averaged URANS equations.

\subsection{Optimization of flow around two-dimensional square cylinder}
\label{sec:Optimization of flow around square cylinder}

So far, the optimization was tested with synthetic reference data created using the forward problem solver and imposing a reference parameter field. This data is ideal in the sense that it was created on the same mesh as the DA is performed. Next, DNS velocity reference data is considered for the flow around a square cylinder. In the following, the test case is described and different reference data point locations are analyzed.

\subsubsection{Test case setup}
\label{sec:Test case setup of flow around two-dimensional square cylinder}

\begin{figure}[!ht]
    \centering
    \includegraphics[]{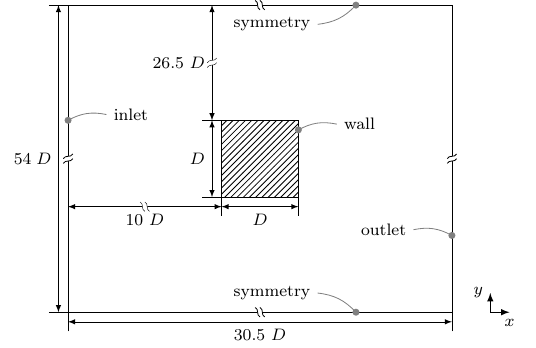}
    \caption{
            Simulation domain of the two-dimensional square cylinder setup. Mean flow is in $x$ direction.
            All length scales are normalized by cylinder width $D$.
    }
    \label{fig:square_2d}
\end{figure}

First, the flow around a two-dimensional square cylinder (e.g.~\cite{trias_turbulent_2015}) is considered. A sketch of the geometry and boundary conditions is provided in Fig.~\ref{fig:square_2d}. All length scales are expressed relative to the cylinder length/height $D$ and the Reynolds number is computed from $D$, the free-stream velocity $u_\infty$ and kinematic viscosity $\nu$. Boundary conditions for velocity and pressure are taken from \cite{trias_turbulent_2015}. Wall functions are applied at the cylinder wall, 2D BC's are present in spanwise directions and symmetry conditions are used for vertical boundaries. At the inlet, Dirichlet BC's are used and set to a small value reproducing the inflow conditions from \cite{trias_turbulent_2015}. Neumann BC's are set at the outlet. In this work, we analyze the case for
\begin{equation*}
    \mathrm{Re}
    =
    \frac{u_{\infty}D}{\nu}
    =
    \num{22000}
\end{equation*}
with DNS velocity reference data from~\cite{trias_turbulent_2015}. A time step size of $\frac{\Delta t u_\infty}{D} = 0.2$ is used.

The mesh was created using \texttt{snappyHexMesh} and consists of 7496 hexahedral cells. The ratio of the smallest to largest cell is $2.487 \cdot 10^{-3}$.

\subsubsection{Reference data in near-cylinder and wake region}
\label{sec:Reference data in near-cylinder and wake region}

A similar reference data point distribution to the one from the circular cylinder setup (see Fig.~\ref{fig:reference data points circular cylinder}) is chosen for the optimization of flow around the square cylinder. Thus, points near the cylinder and in the cylinder wake are chosen where reference data is taken into account for the assimilation. As can be seen in Fig.~\ref{fig:reference data points square cylinder, near cylinder and wake ref}, the density of reference points is higher in the near-cylinder region compared to the wake region. This is due to the more complex flow near the cylinder, which comprises flow separation with high velocity gradients. The cost function depicted in Fig.~\ref{fig:cost function square cylinder, near cylinder and wake ref} reaches its minimum after approximately 100 optimization steps. The initial peaks in the cost function often are appearing when using the \emph{Demon Adam} optimizer and can be explained by its calculation of the moments, which depend on the step size and the maximum number of optimization steps.

\begin{figure}[!ht]
     \centering
     \begin{subfigure}[t]{0.49\textwidth}
         \centering
         \includegraphics[width=\textwidth]{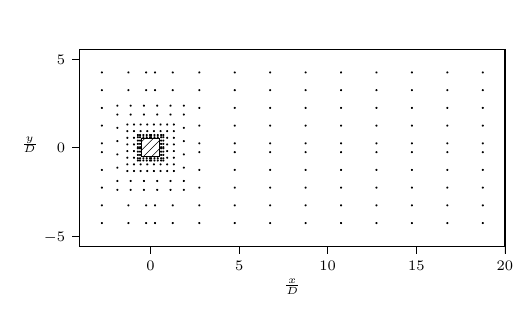}
         \caption{Reference data point locations around square cylinder and its wake. In total 260 points are selected.}
         \label{fig:reference data points square cylinder, near cylinder and wake ref}
     \end{subfigure}
     \hfill
     \begin{subfigure}[t]{0.49\textwidth}
         \centering
         \includegraphics[width=\textwidth]{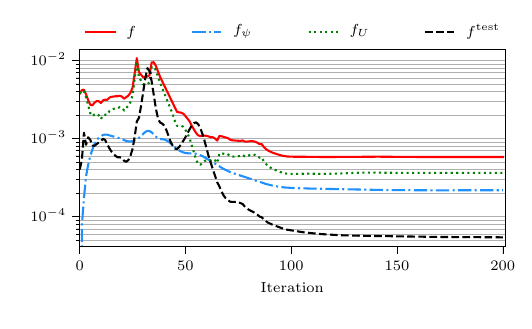}
         \caption{Cost function.}
         \label{fig:cost function square cylinder, near cylinder and wake ref}
     \end{subfigure}
     \hfill
     \begin{subfigure}[b]{0.49\textwidth}
         \centering
         \includegraphics[width=\textwidth]{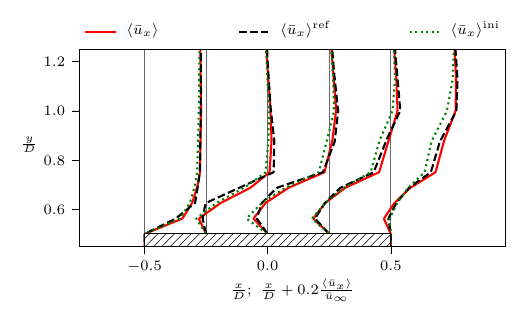}
         \caption{Mean flow velocity profiles near the square cylinder.}
         \label{fig:velocity profiles square cylinder near cylinder, near cylinder and wake ref}
     \end{subfigure}
     \hfill
     \begin{subfigure}[b]{0.49\textwidth}
         \centering
         \includegraphics[width=\textwidth]{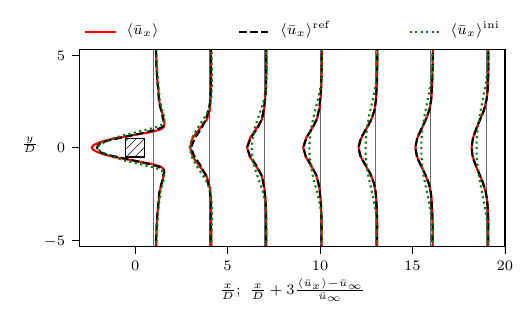}
         \caption{Mean flow velocity profiles in the wake of the square cylinder.}
         \label{fig:velocity profiles square cylinder wake, near cylinder and wake ref}
     \end{subfigure}
        \caption{Optimization of the flow around the two-dimensional square cylinder using DNS reference data from~\cite{trias_turbulent_2015}. Optimization step size is $\eta=\num{7e-5}$ with a maximum number of optimization steps $N=200$. Convergence of the forward problem solver is reached after $C_U=0.5$. The regularization weight parameter is set to $C^{\mathrm{reg}}=\num{7e-3}$.}
        \label{fig:square cylinder, near cylinder and wake ref}
\end{figure}

The velocity profiles near the cylinder as illustrated in Fig.~\ref{fig:velocity profiles square cylinder near cylinder, near cylinder and wake ref} show an improvement toward the reference profiles. However, no perfect match could be achieved. It should be noted though that the resolution near the cylinder wall is quite coarse and hence there are less degrees of freedom for optimization. In addition, this explains the jagged velocity profiles. Fig.~\ref{fig:velocity profiles square cylinder wake, near cylinder and wake ref} depicts that almost all profiles are very well recovered except for the one closest to the cylinder. One reason is that here the density of reference data points switches quite abruptly. Due to the TV regularization all profiles are very smooth, but it should be noted that it also restricts the optimization. Since the regularization weight has to be chosen quite large to yield smooth fields, the optimization does not result in perfect agreement with the reference data in all regions. Generally, these results demonstrate the basic capabilities of our approach and also that the time-averaged DNS velocity data can be well matched.

The baseline solution using the default forward problem solver obtained a Strouhal number of $St=0.095$ for the flow around the square cylinder. The DNS from~\cite{trias_turbulent_2015}, however, relates to $St=0.132$. After the assimilation of time- and spanwise-averaged DNS velocity data, the vortex shedding frequency also adapted, leading to $St=0.12$. This indicates that also with DNS velocity reference data the vortex shedding frequency can be significantly improved.

\subsubsection{Reference data in wake region}
\label{sec:Reference data in wake region}

Up to here, reference data points were distributed around the cylinder and in the wake region. To study the effects of reference data point placement, only points in the cylinder wake are considered. Therefore, a reference data point distribution as shown in Fig.~\ref{fig:reference data points square cylinder, wake ref} is chosen, which is almost the same as the previous distribution from Fig.~\ref{fig:reference data points square cylinder, near cylinder and wake ref} without the near-cylinder points. As before, the discrepancy function $f_U$ decreases at least by one order of magnitude and also the test function minimizes. Since the data points are quite sparsely (every $2D$ in $x$ direction and every $D$ in $y$ direction) distributed, a higher regularization weight is needed. Hence, the resulting overall cost function $f$ still is quite large (cf.~Fig.~\ref{fig:cost function square cylinder, wake ref}).

The velocity profiles in the cylinder wake match to a similar degree with the reference as observed before (compare~Fig.~\ref{fig:velocity profiles square cylinder wake, near cylinder and wake ref} and Fig.~\ref{fig:velocity profiles square cylinder wake, wake ref}). The velocity peaks at the center line ($\nicefrac{y}{D}=0$), however, show a little offset from the reference velocity peaks. When looking at the velocity profiles near the cylinder (see Fig.~\ref{fig:velocity profiles square cylinder near cylinder, wake ref}), it is apparent that no improvement took place. Therefore, the optimization parameter or divergence-free force in conjunction with the TV regularization has not enough impact further upstream of $\nicefrac{x}{D} \approx 2.5$ to also reconstruct the mean flow in the near-cylinder region. In Section~\ref{sec:Comparison of cylinder divergence} it further is elaborated on the differences between stationary divergence-free forcing terms when using different reference data point distributions.

\begin{figure}[!ht]
     \centering
     \begin{subfigure}[t]{0.49\textwidth}
         \centering
         \includegraphics[width=\textwidth]{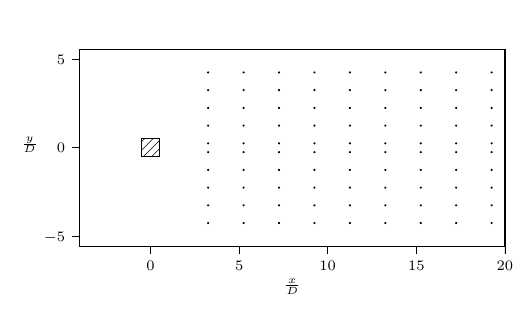}
         \caption{Reference data point locations around square cylinder and its wake. In total 90 points are selected.}
         \label{fig:reference data points square cylinder, wake ref}
     \end{subfigure}
     \hfill
     \begin{subfigure}[t]{0.49\textwidth}
         \centering
         \includegraphics[width=\textwidth]{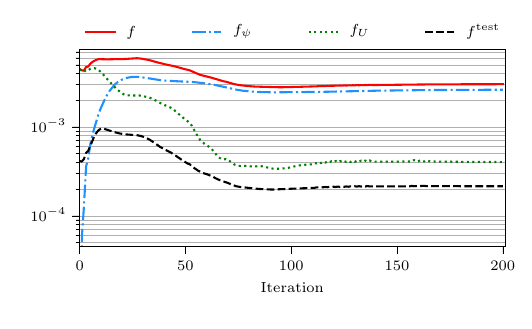}
         \caption{Cost function.}
         \label{fig:cost function square cylinder, wake ref}
     \end{subfigure}
     \hfill
     \begin{subfigure}[b]{0.49\textwidth}
         \centering
         \includegraphics[width=\textwidth]{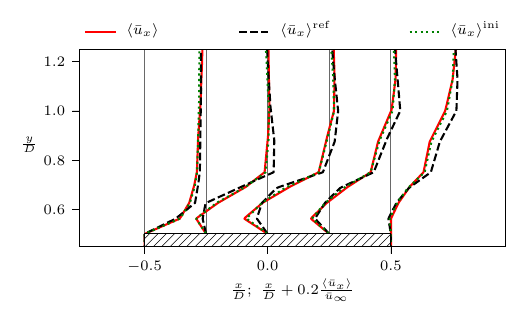}
         \caption{Mean flow velocity profiles near the square cylinder.}
         \label{fig:velocity profiles square cylinder near cylinder, wake ref}
     \end{subfigure}
     \hfill
     \begin{subfigure}[b]{0.49\textwidth}
         \centering
         \includegraphics[width=\textwidth]{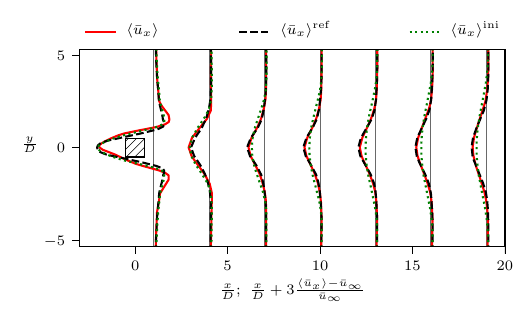}
         \caption{Mean flow velocity profiles in the wake of the circular cylinder.}
         \label{fig:velocity profiles square cylinder wake, wake ref}
     \end{subfigure}
        \caption{Optimization of flow around the two-dimensional square cylinder using DNS reference data from~\cite{trias_turbulent_2015}. Optimization step size is $\eta=\num{2e-5}$ with a maximum number of optimization steps $N=200$. Convergence of the forward problem solver is reached after $C_U=0.5$. The regularization weight parameter is set to $C^{\mathrm{reg}}=\num{7e-2}$.}
        \label{fig:square cylinder, wake ref}
\end{figure}

After the assimilation of time- and spanwise-averaged DNS velocity data, the Strouhal number stayed unchanged at $St=0.095$. Thus, the assimilation of DNS velocity data in the cylinder wake does not suffice for recovering the vortex shedding frequency from the DNS reference, but allows for mean flow reconstruction in the wake region. 

\subsubsection{Reference data in near-cylinder region}
\label{sec:Reference data in near-cylinder region}

Since the placement of reference points in the cylinder wake did not improve the dynamics (Strouhal number), the influence of reference points close to the cylinder is examined in this section. For that, a reference data point distribution as presented in Fig.~\ref{fig:reference data points square cylinder, near cylinder ref} is adopted. As can be seen in Fig.~\ref{fig:cost function square cylinder, near cylinder ref}, the optimization works as observed the times before, that is, discrepancy and test functions are decreasing. 

\begin{figure}[!ht]
     \centering
     \begin{subfigure}[t]{0.49\textwidth}
         \centering
         \includegraphics[width=\textwidth]{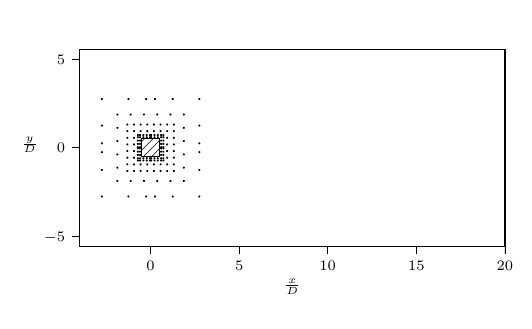}
         \caption{Reference data point locations around square cylinder and its wake. In total 152 points are selected.}
         \label{fig:reference data points square cylinder, near cylinder ref}
     \end{subfigure}
     \hfill
     \begin{subfigure}[t]{0.49\textwidth}
         \centering
         \includegraphics[width=\textwidth]{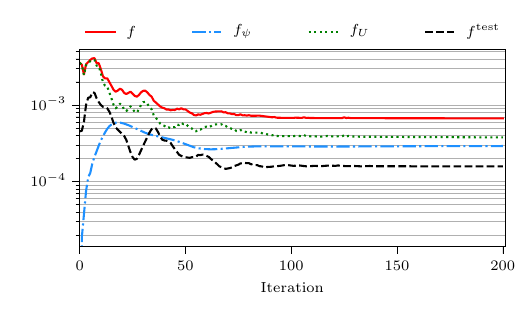}
         \caption{Cost function.}
         \label{fig:cost function square cylinder, near cylinder ref}
     \end{subfigure}
     \hfill
     \begin{subfigure}[b]{0.49\textwidth}
         \centering
         \includegraphics[width=\textwidth]{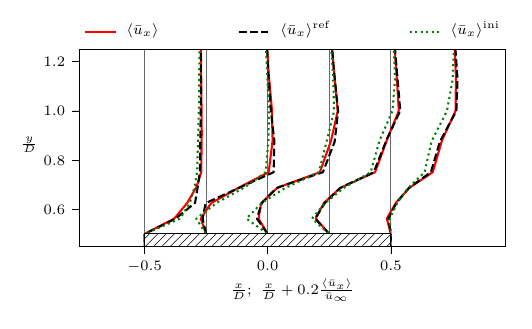}
         \caption{Mean flow velocity profiles near the square cylinder.}
         \label{fig:velocity profiles square cylinder near cylinder, near cylinder ref}
     \end{subfigure}
     \hfill
     \begin{subfigure}[b]{0.49\textwidth}
         \centering
         \includegraphics[width=\textwidth]{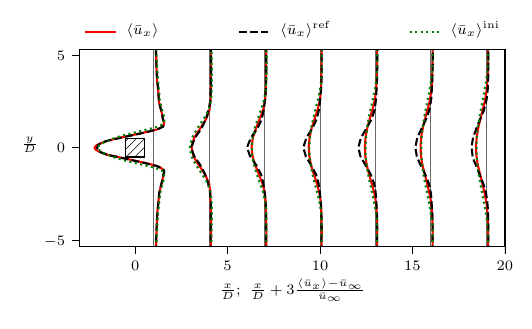}
         \caption{Mean flow velocity profiles in the wake of the square cylinder.}
         \label{fig:velocity profiles square cylinder wake, near cylinder ref}
     \end{subfigure}
        \caption{Optimization of flow around the two-dimensional square cylinder using DNS reference data from~\cite{trias_turbulent_2015}. Optimization step size is $\eta=\num{7e-5}$ with a maximum number of optimization steps $N=200$. Convergence of the forward problem solver is reached after $C_U=0.5$. The regularization weight parameter is set to $C^{\mathrm{reg}}=\num{2e-3}$.}
        \label{fig:square cylinder, near cylinder ref}
\end{figure}

Again, thanks to regularization smooth velocity profiles (see Fig.~\ref{fig:velocity profiles square cylinder near cylinder, near cylinder ref} and Fig.~\ref{fig:velocity profiles square cylinder wake, near cylinder ref}) are obtained. Since only reference points near the cylinder are considered, velocity profiles in the near-wake improve toward the reference, but in the far-wake region (cf.~Fig.~\ref{fig:velocity profiles square cylinder wake, near cylinder ref}) they stay almost unchanged. The velocity profiles close to the cylinder wall, however, show a compelling enhancement and match very well with the reference velocity profiles. Compared to the results obtained from using the reference data point distribution that considers both near-cylinder points and wake points, the velocity profiles also get corrected in the very first cells close to the wall. Since the regularization weight for these two regions with different spacing of the reference data points are ideally not the same, the regularization might hinder the optimization too much in some of the regions. 

The very good match between optimized and reference mean velocity profiles correlates well with the improvement of the dynamic behavior of the flow. The Strouhal number changed from $St=0.095$ to $St=0.122$, nearing the DNS reference ($St=0.132$). Consequently, near-cylinder reference points play a crucial role in recovering the flow dynamics, i.e., the vortex shedding expressed through the Strouhal number improved. However, if mean flow reconstruction is the major objective, near-cylinder reference points alone do not suffice and reference points in the cylinder wake are also needed.

\subsubsection{Comparison of divergence-free forcing fields}
\label{sec:Comparison of cylinder divergence}

\begin{figure}[!ht]
     \centering
     \begin{subfigure}[t]{0.32\textwidth}
         \centering
         \includegraphics[width=\textwidth]{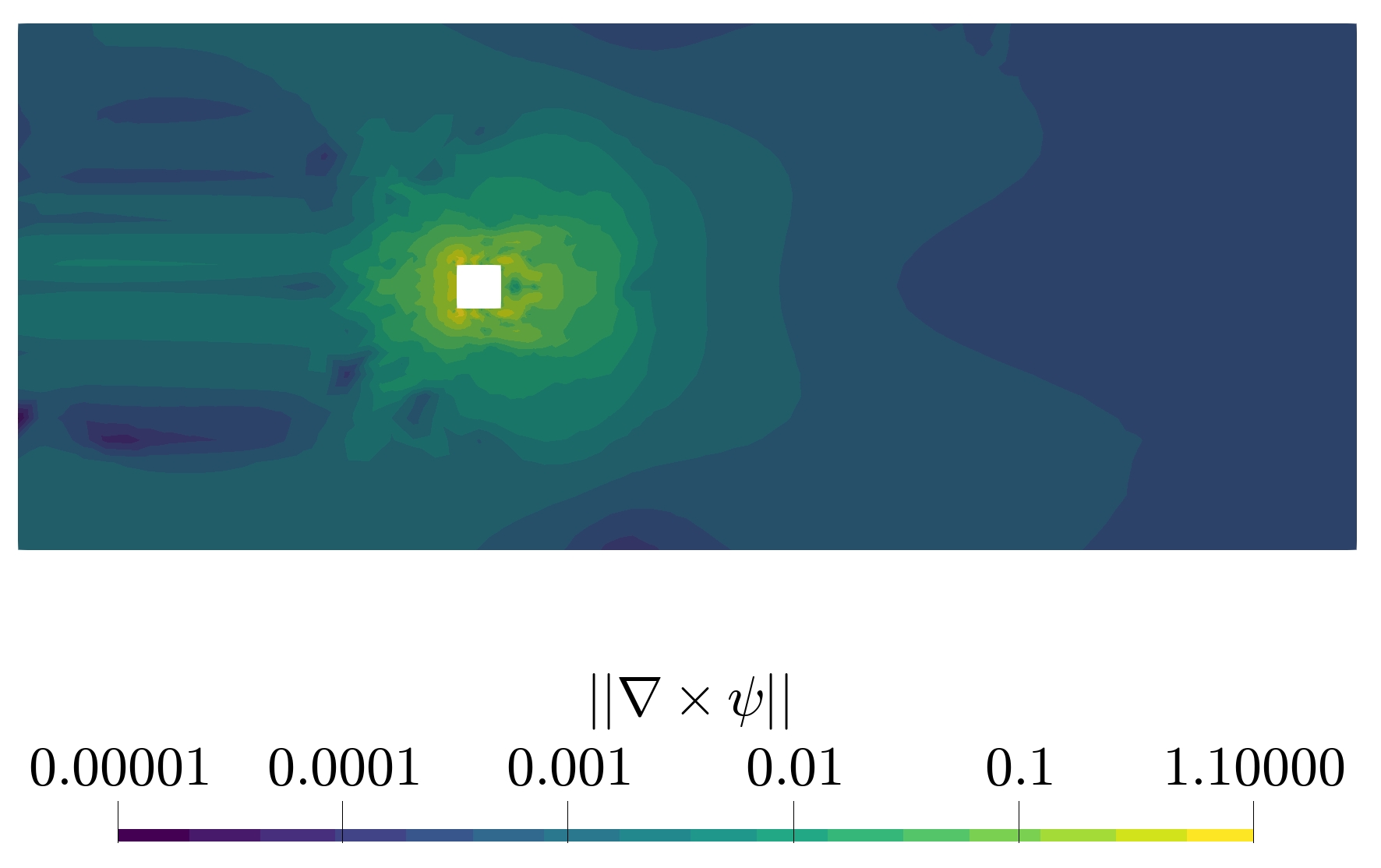}
         \caption{Near-cylinder reference data points.}
         \label{fig:curl psi near cylinder}
     \end{subfigure}
     \hfill
     \begin{subfigure}[t]{0.32\textwidth}
         \centering
         \includegraphics[width=\textwidth]{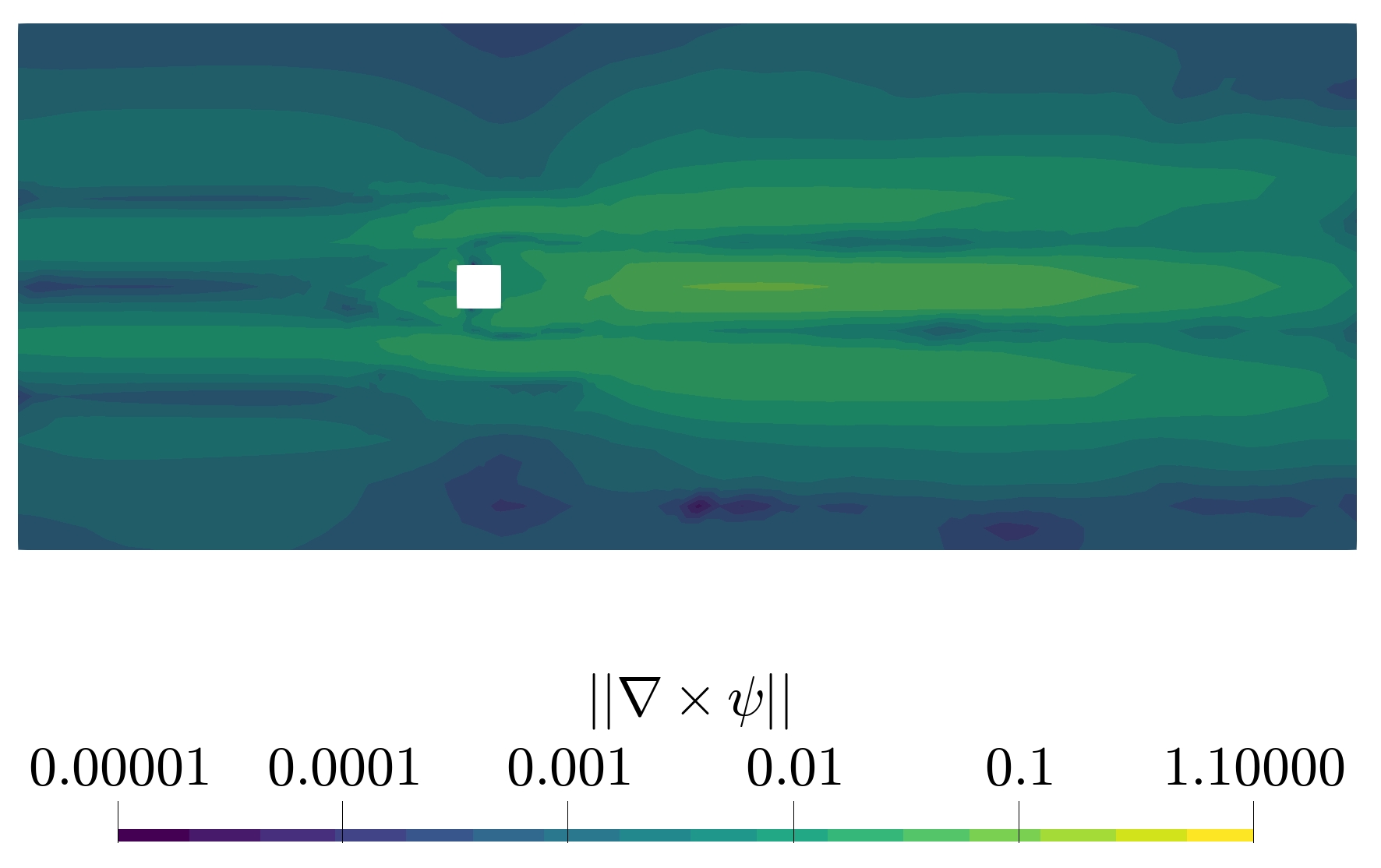}
         \caption{Wake reference data points.}
         \label{fig:curl psi wake}
     \end{subfigure}
     \hfill
     \begin{subfigure}[t]{0.32\textwidth}
         \centering
         \includegraphics[width=\textwidth]{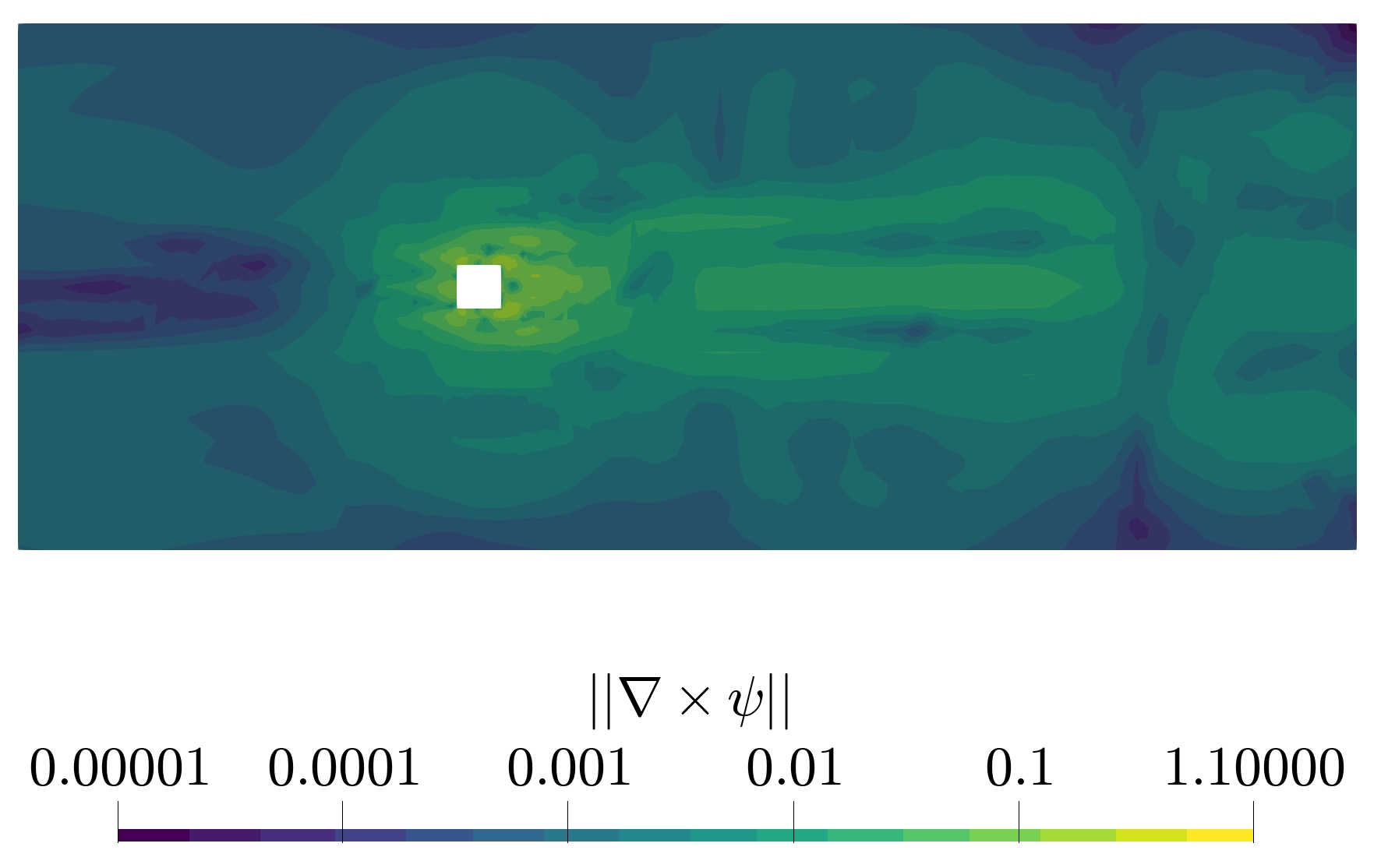}
         \caption{Near-cylinder and wake reference data points.}
         \label{fig:curl psi near cylinder wake}
     \end{subfigure}
        \caption{Comparison of divergence-free forcing $\lVert \nabla \times \psi \rVert$ for different reference data point locations taken after the optimization is converged, i.e., after the 200th optimization step.}
        \label{fig:curl psi comparison}
\end{figure}

Earlier results demonstrate that the choice of reference data points distribution heavily influences the region where the mean flow velocity is reconstructed, and also if the prediction of the flow dynamics, i.e., of the vortex shedding frequency improved. The frequency of the vortex shedding only can change since the stationary divergence-free forcing term $\nabla \times \psi$ acts in the URANS equations of the forward problem solver during each optimization step and updates according to the solution of the inverse problem. The corrective forcing term $\epsilon_{ijk} \frac{\partial \psi_k}{\partial x_j}$, as presented in Section~\ref{sec:Data assimilation parameter}, aims to correct the anisotropic part of the Reynolds stress tensor. One could write
\begin{equation}
    - \frac{\partial a_{ij}^\mathrm{corr}}{\partial x_j}
    =
    \underbrace{\epsilon_{ijk} \frac{\partial \psi_k}{\partial x_j}}_{f_i^\mathrm{corr}} 
    \ ,
\end{equation}
where $a_{ij}^\mathrm{corr}$ is the correction of the anisotropic part of the Reynolds stress tensor and $f_i^\mathrm{corr}$ the discussed divergence-free forcing term. Thus, the corrective forcing term accounts for the discrepancies in the modeled Reynolds stresses from the eddy viscosity model. Consequently, for the three different distributions of reference data points, the divergence-free forcing term $\nabla \times \psi$ is depicted in Fig.~\ref{fig:curl psi comparison}. Firstly, it is apparent that for the near-cylinder reference data distribution (see Fig.~\ref{fig:curl psi near cylinder}), the forcing shows the highest magnitude out of all three cases. This peak is located right next to the cylinder wall, which explains the great optimization in that region. However, as expected from previous analysis, the forcing decays quickly further downstream. Hence, no improvement in the velocity field for the far-wake region could be achieved. Secondly, Fig.~\ref{fig:curl psi wake} depicts the wider spatial range of the forcing where only wake points were considered. Interestingly, the forcing prevails also upstream of the cylinder. Nevertheless, the magnitude is much smaller and especially near the cylinder wall it has much less impact. Lastly, Fig.~\ref{fig:curl psi near cylinder wake} illustrates the forcing behavior for reference points near the cylinder and in the wake. As somehow expected, the spatial contours of the forcing combine the aspects of the two other reference point distributions. Therefore, a strong influence is evident in both regions. Nonetheless, the magnitude is smaller than for the forcing using only near-cylinder reference points (compare Fig.~\ref{fig:curl psi near cylinder} and Fig.~\ref{fig:curl psi near cylinder wake}). Moreover, this explains the discrepancies in the degree of optimization in the near-wall region of the cylinder (cf. Fig.~\ref{fig:velocity profiles square cylinder near cylinder, near cylinder and wake ref} and Fig.~\ref{fig:velocity profiles square cylinder near cylinder, near cylinder ref}).

\subsection{Optimization of flow around two-dimensional rectangular cylinder}
\label{sec:Optimization of flow around rectangular cylinder}

\subsubsection{Test case setup}
\label{sec:Test case setup of flow around two-dimensional rectangular cylinder}

\begin{figure}[!ht]
    \centering
    \includegraphics[]{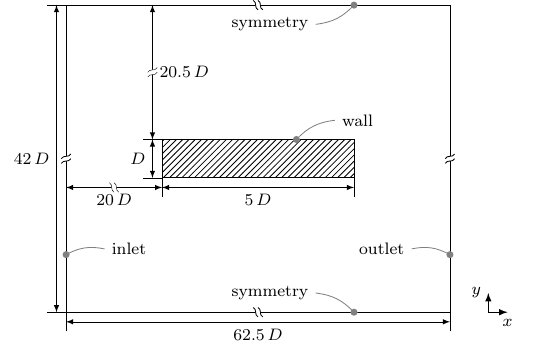}
    \caption{
            Simulation domain of the two-dimensional rectangular cylinder (BARC) setup with a chord-to-thickness ratio of 5:1. Mean flow is in $x$ direction.
            All length scales are normalized by cylinder thickness $D$.
    }
    \label{fig:rectangular_cylinder_2d}
\end{figure}

Lastly, the flow around a two-dimensional rectangular cylinder (e.g.~\cite{chiarini_turbulent_2021}) is considered. A sketch of the geometry and boundary conditions is provided in Fig.~\ref{fig:rectangular_cylinder_2d}. All length scales are expressed relative to the cylinder height $D$ and the Reynolds number is computed from $D$, the free-stream velocity $u_\infty$ and kinematic viscosity $\nu$. Boundary conditions for velocity and pressure are taken from \cite{chiarini_turbulent_2021}. Wall functions are applied at the cylinder wall, 2D BC's are present in spanwise directions and symmetry conditions are used for vertical boundaries. At the inlet, Dirichlet BC's are used and set to a small value reproducing the inflow conditions from \cite{chiarini_turbulent_2021}. Neumann BC's are set at the outlet. Flow conditions at a Reynolds number of
\begin{equation*}
    \mathrm{Re}
    =
    \frac{u_{\infty}D}{\nu}
    =
    \num{3000}
\end{equation*}
are considered together with DNS velocity reference data from~\cite{chiarini_turbulent_2021}. A time step size of $\frac{\Delta t u_\infty}{D} = 0.3$ is used.

The mesh was created using \texttt{snappyHexMesh} and consists of 11062 hexahedral cells. The ratio of the smallest to largest cell is $3.108 \cdot 10^{-3}$.

\subsubsection{Reference data in near-cylinder and wake region}
\label{sec:Reference data in near-cylinder and wake region, rectangular cylinder}

As discussed in Section~\ref{sec:Optimization of flow around square cylinder}, reference data points should be placed in both the near-cylinder as well as in the wake regions to benefit from mean flow reconstruction and partial dynamic flow improvement. Therefore, only this reference data configuration is presented and elaborated on in this section. Due to a chord-to-thickness ratio of 5:1, more reference data points were allocated close to the cylinder (see Fig.~\ref{fig:reference data points rectangular cylinder, near cylinder and wake ref}). The optimization results in the typical behavior for cost and test functions (cf.~Fig.~\ref{fig:cost function rectangular cylinder, near cylinder and wake ref}) as seen before rooting from the \emph{Demon Adam} optimization algorithm.

\begin{figure}[!ht]
     \centering
     \begin{subfigure}[t]{0.49\textwidth}
         \centering
         \includegraphics[width=\textwidth]{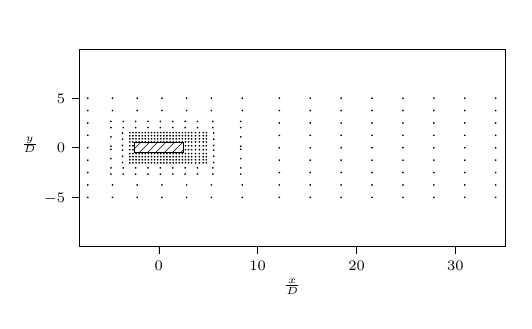}
         \caption{Reference data point locations around rectangular cylinder and its wake. In total 392 points are selected.}
         \label{fig:reference data points rectangular cylinder, near cylinder and wake ref}
     \end{subfigure}
     \hfill
     \begin{subfigure}[t]{0.49\textwidth}
         \centering
         \includegraphics[width=\textwidth]{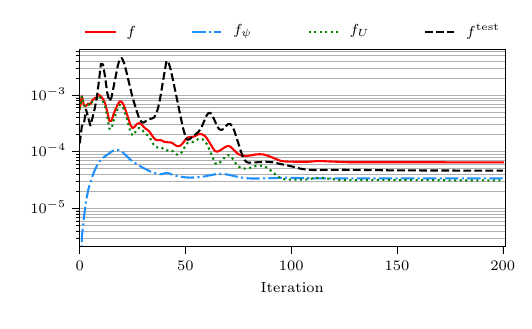}
         \caption{Cost function.}
         \label{fig:cost function rectangular cylinder, near cylinder and wake ref}
     \end{subfigure}
     \hfill
     \begin{subfigure}[b]{0.49\textwidth}
         \centering
         \includegraphics[width=\textwidth]{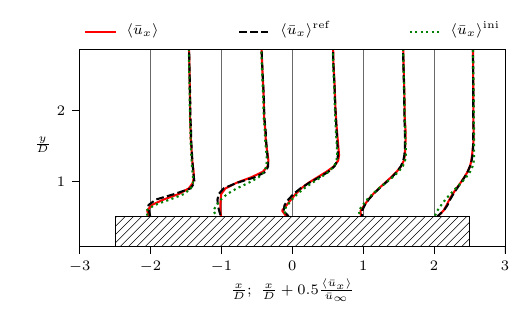}
         \caption{Mean flow velocity profiles near the rectangular cylinder.}
         \label{fig:velocity profiles rectangular cylinder near cylinder, near cylinder and wake ref}
     \end{subfigure}
     \hfill
     \begin{subfigure}[b]{0.49\textwidth}
         \centering
         \includegraphics[width=\textwidth]{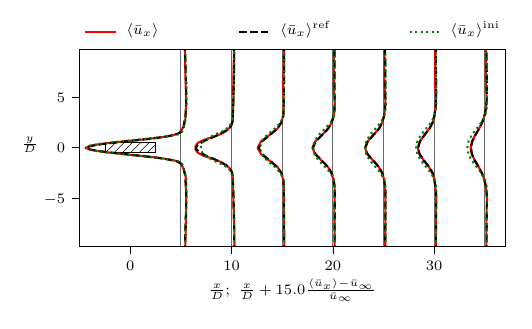}
         \caption{Mean flow velocity profiles in the wake of the rectangular cylinder.}
         \label{fig:velocity profiles rectangular cylinder wake, near cylinder and wake ref}
     \end{subfigure}
        \caption{Optimization of flow around the two-dimensional rectangular cylinder using DNS reference data from~\cite{chiarini_turbulent_2021}. Optimization step size is $\eta=\num{3e-5}$ with a maximum number of optimization steps $N=200$. Convergence of the forward problem solver is reached after $C_U=0.5$. The regularization weight parameter is set to $C^{\mathrm{reg}}=\num{6e-4}$.}
        \label{fig:rectangular cylinder, near cylinder and wake ref}
\end{figure}

The baseline solution already provides a very good initial guess indicated by the small discrepancy in the velocity profiles as depicted in Fig.~\ref{fig:velocity profiles rectangular cylinder near cylinder, near cylinder and wake ref} and Fig.~\ref{fig:velocity profiles rectangular cylinder wake, near cylinder and wake ref} for near-cylinder and wake regions, respectively. In conjunction with the decrease of the discrepancy part of the cost function as well as the test function, the velocity profiles also get corrected toward the reference velocity profiles. Again, this improvement projects onto the flow dynamics, as the Strouhal number adjusts from $St=0.0944$ (baseline solution) to $St=0.1111$. The DNS reference \cite{chiarini_turbulent_2021} states $St=0.1274$, and thus also in this case assimilation of time-averaged data allowed a partial recovery of the flow dynamics.


\section{Conclusions and outlook}
\label{sec:conclusion}
We argue that URANS simulations, despite their drawbacks, still are of great importance for the analysis of complex and industrial flow problems that involve severe unsteadiness, e.g. vortex shedding. The accuracy of these models can be improved by various approaches. The goal of this work was to perform variational data assimilation of sparse time-averaged data into an URANS simulation by means of a stationary divergence-free forcing term in the URANS momentum equations. A stationary discrete adjoint approach was proposed to compute the cost function gradient at low computational cost. To do so, the URANS equations were time-averaged during their solution (forward problem) and subsequently used to construct a stationary adjoint equation (inverse problem). Making use of our efficient semi-analytical approach to compute the cost function gradient and the gradient-based \textit{Demon Adam} optimizer for the parameter update, we demonstrated that for coarse URANS simulations the mean velocity profiles match well with the reference. Since the updated parameter acts in the instantaneous URANS equations, an improvement of the dynamic flow behavior could be observed for all flow cases that were investigated. That is, after optimization the Strouhal number that characterizes the vortex shedding is in better agreement with the Strouhal number from the high-fidelity DNS than it was before. However, the obtained results considerably depend on the location of the mean velocity reference points. Our findings show that the near-cylinder region is crucial to improve the vortex shedding frequency, while data points further downstream are necessary, if a corrected time-averaged velocity field in the wake region is of interest. 

Future work will focus on extending this framework toward the use of weighted time-averages, that is, not only considering one time-average, but multiple averages weighted to varying degrees of importance of times in the evolution of the time signal. Additionally, the placement of reference data points should further be investigated to reduce the number of necessary observations, while respecting local restrictions (e.g. wake and near-cylinder points) for an optimal mean flow reconstruction and dynamic flow control. For example, Franz~\etal.~\cite{franz22} employed dynamic reference data patterns during the DA process, which drastically improved the convergence rates.

We would especially like to point out that the presented adjoint-based DA framework is not limited to the use of URANS models, but LES models are also promising candidates. Analogously to the presented URANS optimizations, the discrepancies between results from LES and a high-fidelity measurement or simulation due to assumptions in the sub-grid scale models, mesh resolution etc. could also be minimized. 

The divergence-free forcing fields obtained by the presented DA approach could also serve as an input for data-driven closure model training and support the ongoing development of eddy viscosity and sub-grid scale models.


\section*{Acknowledgments}
\label{sec:acknowledgments}
The authors would like to thank Robert Epp and Pasha Piroozmand for fruitful discussions. We also appreciate Alessandro Chiarini and Maurizio Quadrio for providing the DNS reference data of the flow around the rectangular cylinder. Financial support from the Swiss Federal Office for Energy SFOE, project ReMOVES2, contract number SI/502085-01 is gratefully acknowledged.


\appendix

\setcounter{figure}{0}    

\section{Temporal averaging of URANS equations}
\label{sec:Temporal averaging of URANS equations appendix}

Apply temporal averaging $\left<\cdot\right>$ to Eq.~\eqref{eq:rans_momentum}:
\begin{align}
    & \cancel{\frac{\partial \left<\bar{u}_{i}\right>}{\partial t}}
    + \frac{\partial \left<\bar{u}_{i}\right> \left<\bar{u}_{j}\right>}{\partial x_{j}} \\
    &= -\frac{\partial \left<\bar{p}\right>}{\partial x_{i}}
    + \frac{\partial}{\partial x_{j}}
    \left<
        2\left(\nu + \nu_{t}\right) \bar{S}_{ij} 
    \right>
    - \frac{\partial \left<\bar{u}''_{i} \bar{u}''_{j}\right>}{\partial x_{j}} 
     + \epsilon_{ijk} \frac{\partial \left< \psi_{k}\right>}{\partial x_{j}}\\
    &= -\frac{\partial \left<\bar{p}\right>}{\partial x_{i}}
    + \frac{\partial}{\partial x_{j}}
    \left(
        2 \nu \left<\bar{S}_{ij} \right>
        +  2 \left<
            \nu_{t} \bar{S}_{ij}
        \right>
    \right)
    - \frac{\partial \left<\bar{u}''_{i} \bar{u}''_{j}\right>}{\partial x_{j}} + \epsilon_{ijk} \frac{\partial \psi_{k}}{\partial x_{j}}\\
        &= -\frac{\partial \left<\bar{p}\right>}{\partial x_{i}}
    + \frac{\partial}{\partial x_{j}}
    \left(
        2 \nu \left<\bar{S}_{ij} \right>
    \right)
    + \frac{\partial}{\partial x_{j}}
    \left(
         2 \left<
            \left(\left<\nu_{t}\right> + \nu_{t}''\right)
            \left(
                \left<\bar{S}_{ij} \right>
                + \bar{S}_{ij}''
            \right)
        \right>
    \right)
    - \frac{\partial \left<\bar{u}''_{i} \bar{u}''_{j}\right>}{\partial x_{j}} + \epsilon_{ijk} \frac{\partial \psi_{k}}{\partial x_{j}}\\
        &= -\frac{\partial \left<\bar{p}\right>}{\partial x_{i}}
    + \frac{\partial}{\partial x_{j}}
    \left(
        2 \left(\nu + \left<\nu_{t}\right>\right) \left<\bar{S}_{ij} \right>
    \right)
    + \frac{\partial}{\partial x_{j}}
    \left(
        2 \left<\nu''_{t} \bar{S}_{ij}'' \right>
    \right)
    - \frac{\partial \left<\bar{u}''_{i} \bar{u}''_{j}\right>}{\partial x_{j}}
    + \epsilon_{ijk} \frac{\partial \psi_{k}}{\partial x_{j}}
\end{align}

\section{Discrete adjoint method}
\label{sec:Discrete adjoint method appendix}

For the solution vector we write
\begin{equation}
\label{eq:solution vector}
    U
    =
    \left[
        \langle\bar{u}_{x}\rangle, \, \langle\bar{u}_{y}\rangle, \, \langle\bar{u}_{z}\rangle, \, \langle\bar{p}\rangle
    \right]^{T}
    \ .
\end{equation}
The stationary forward system of equations, i.\,e. the time-averaged URANS equations are formulated in a coupled manner:
\begin{equation}
    \label{eq:coupled_residual}
    R\left(\psi, U\right)
    =
    \begin{bmatrix}
        R_{\langle\bar{u}\rangle}\left(\psi, U\right) \\
        R_{\langle\bar{p}\rangle}\left(\psi, U\right)
    \end{bmatrix}
    =
    \begin{bmatrix}
        \frac{\partial R_{\langle\bar{u}\rangle}}{\partial \langle\bar{u}\rangle} & \frac{\partial R_{\langle\bar{u}\rangle}}{\partial \langle\bar{p}\rangle} \\
        \frac{\partial R_{\langle\bar{p}\rangle}}{\partial \langle\bar{u}\rangle} & \frac{\partial R_{\langle\bar{p}\rangle}}{\partial \langle\bar{p}\rangle}
    \end{bmatrix}
    \begin{bmatrix}
        \langle\bar{u}\rangle \\
        \langle\bar{p}\rangle
    \end{bmatrix}
    -
    \begin{bmatrix}
        b_{\langle\bar{u}\rangle} \\
        b_{\langle\bar{p}\rangle}
    \end{bmatrix}
    =
    \mathbf{A}_{U}
    U
    -
    b_{U}
    =
    0
    \ .
\end{equation}

To this end, we introduce a Lagrangian
\begin{equation*}
    \mathcal{L}\left(\psi, U\right)
    =
    f\left(\psi, U\right)
    -
    \lambda^{T} R\left(\psi, U\right)
\end{equation*}
with Lagrange multipliers
\begin{equation}
    \lambda
    =
    \left[
        \lambda_{\left<\bar{u}_{x}\right>}, \, \lambda_{\left<\bar{u}_{y}\right>}, \, \lambda_{\left<\bar{u}_{z}\right>}, \, \lambda_{\left<\bar{p}\right>}
    \right]^{T}
    \ .
\end{equation}

The respective gradient with respect to the parameters is derived as
\begin{align}
    \label{eq:adjoint_gradient}
    \frac{\mathrm{d} f}{\mathrm{d} \psi}
    &=
    \nonumber
    \frac{\mathrm{d} \mathcal{L}}{\mathrm{d} \psi} \\
    \nonumber
    &=
    \frac{\partial f}{\partial \psi}
    +
    \frac{\partial f}{\partial U} \frac{\partial U}{\partial \psi}
    -
    \lambda^{T} \frac{\partial R}{\partial \psi}
    -
    \lambda^{T} \frac{\partial R}{\partial U} \frac{\partial U}{\partial \psi} \\
    \nonumber
    &=
    \frac{\partial f_{\psi}}{\partial \psi}
    +
    \frac{\partial f_{U}}{\partial U} \frac{\partial U}{\partial \psi}
    -
    \lambda^{T} \frac{\partial R}{\partial \psi}
    -
    \lambda^{T} \frac{\partial R}{\partial U} \frac{\partial U}{\partial \psi} \\
    \nonumber
    &=
    \frac{\partial f_{\psi}}{\partial \psi}
    -
    \lambda^{T} \frac{\partial R}{\partial \psi}
    +
    \Biggl[
        \underbrace{
            \frac{\partial f_{U}}{\partial U}
            - \lambda^{T} \frac{\partial R}{\partial U}
        }_{=0}
    \Biggr]
    \frac{\partial U}{\partial \psi} \\
    &=
    \frac{\partial f_{\psi}}{\partial \psi}
    -
    \lambda^{T} \frac{\partial R}{\partial \psi}
    \ .
\end{align}
The degrees of freedom introduced by the Lagrange multipliers $\lambda$ are used to demand the expression in brackets to vanish.

This yields the following equation for $\lambda$:
\begin{equation}
    \label{eq:coupled_adjoint_appendix}
    \left( \frac{\partial R}{\partial U} \right)^{T}
    \lambda
    =
    \mathbf{A}_{U}^T \lambda 
    =
    \left( \frac{\partial f_{U}}{\partial U} \right)^{T} .
\end{equation}
where $\mathbf{A}_{U}$ is the stationary forward system matrix. For more details please refer to Brenner~\etal.~\cite{brenner22}.

Applied to the time-averaged URANS forward problem with residual~\eqref{eq:coupled_residual}, the \emph{coupled} adjoint system of equations in Eq.~\eqref{eq:coupled_adjoint} reads
\begin{equation}
    \label{eq:coupled_adjoint_rans}
    \renewcommand\arraystretch{1.2}
    \begin{bmatrix}
        \left(\frac{\partial R_{\left<\bar{u}\right>}}{\partial \left<\bar{u} \right>}\right)^{T} & \left(\frac{\partial R_{\left<\bar{p}\right>}}{\partial \left<\bar{u}\right>}\right)^{T} \\
        \left(\frac{\partial R_{\left<\bar{u}\right>}}{\partial \left<\bar{p}\right>}\right)^{T} & \left(\frac{\partial R_{\left<\bar{p}\right>}}{\partial \left<\bar{p}\right>}\right)^{T}
    \end{bmatrix}
    \begin{bmatrix}
        \lambda_{\left<\bar{u}\right>} \\
        \lambda_{\left<\bar{p}\right>}
    \end{bmatrix}
    =
    \begin{bmatrix}
        \left( \frac{\partial f_{\left<\bar{u}\right>}}{\partial \left<\bar{u}\right>} \right)^{T} \\
        \left( \frac{\partial f_{\left<\bar{p}\right>}}{\partial \left<\bar{p}\right>} \right)^{T}
    \end{bmatrix}
    =
    \begin{bmatrix}
        \left( \frac{\partial f_{\left<\bar{u}\right>}}{\partial \left<\bar{u}\right>} \right)^{T} \\
        0
    \end{bmatrix}
    ,
\end{equation}

since only velocity data is assimilated. Parameter $\psi$ only appears in a single term in $R_{\left<\bar{u}\right>}$ (cf. Eq.~\eqref{eq:urans_time_average}), i.\,e.,
\begin{equation}
    \label{eq:dRdPsi}
    \frac{\partial R_{\left<\bar{u}\right>}}{\partial \psi}
    =
    \frac{\partial}{\partial \psi}
    \left[
        -\epsilon_{ijk} \frac{\partial \psi_{k}}{\partial x_{j}}
    \right]
    \approx -\frac{\partial}{\partial \psi}
    \left[
        \mathbf{A}_{\psi} \psi - b_{\psi}
    \right]
    =
    -\mathbf{A}_{\psi}
    \ ,
\end{equation}
where $\mathbf{A}_{\psi}$ is the system matrix and $b_{\psi}$ the right hand side vector of the implicit curl operator, respectively.

\section{Synthetic reference data generation}
\label{sec:Synthetic reference data generation}

\begin{figure}[!ht]
    \centering
    \includegraphics[width=0.3\textwidth]{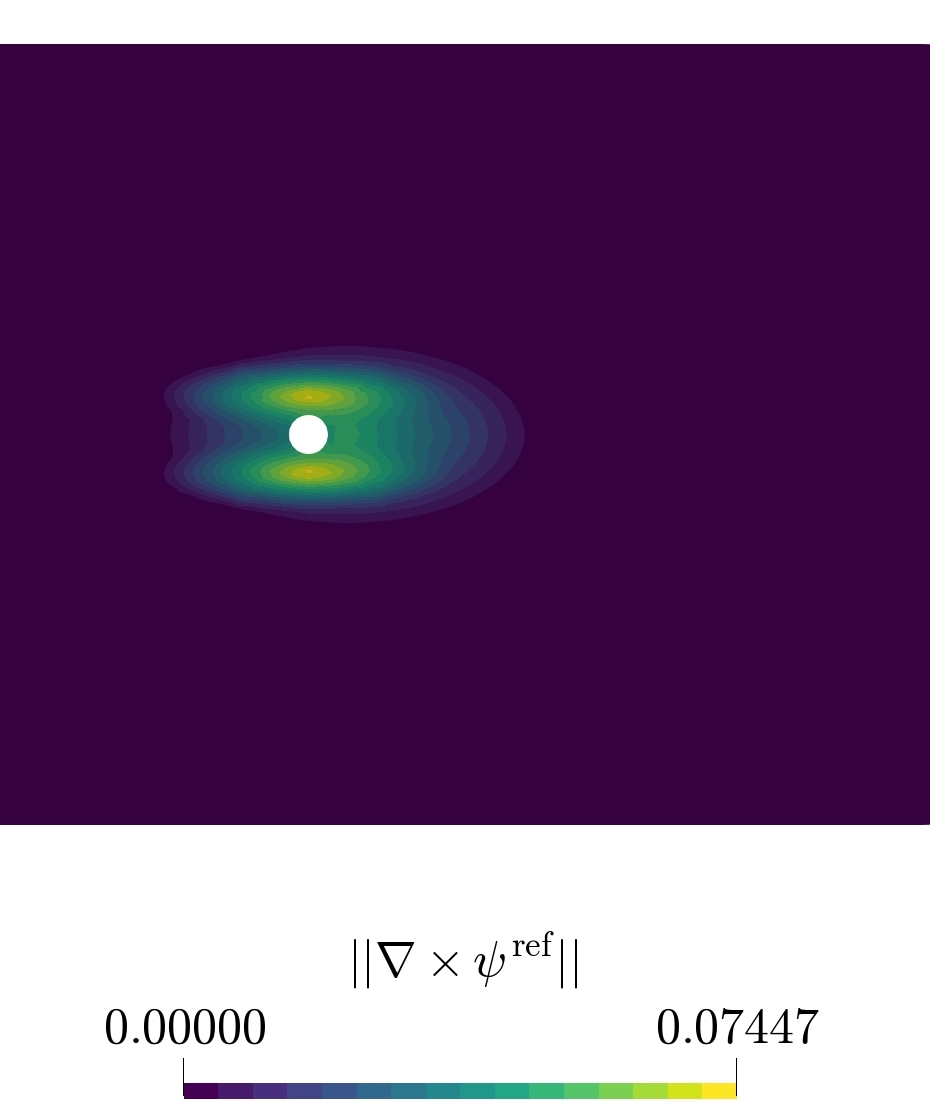}
    \caption{
            Creation of synthetic reference using the forward solver and an imposed divergence-free forcing vector $\nabla \times \psi^\mathrm{ref}$.
    }
    \label{fig:synthetic ref curl psi creation}
\end{figure}

\begin{figure}[!ht]
    \centering
    \includegraphics[width=0.3\textwidth]{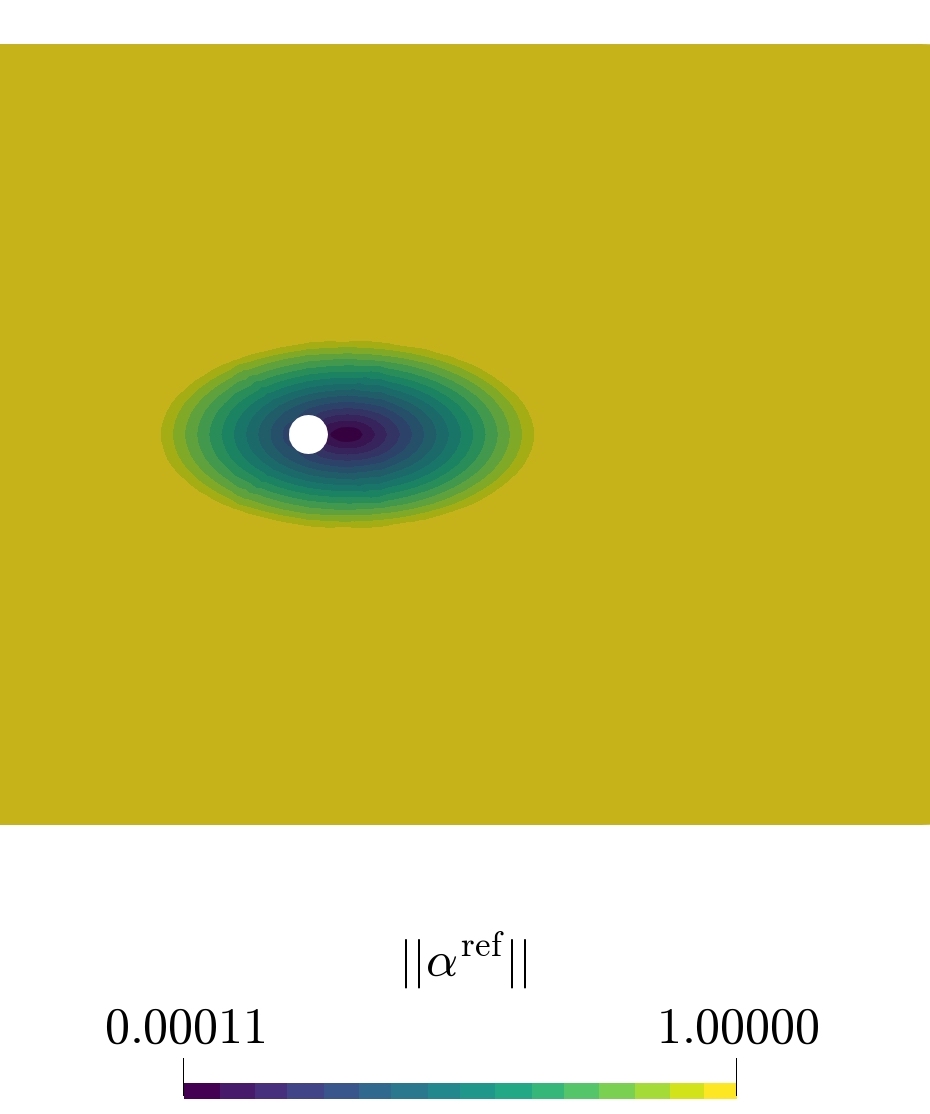}
    \caption{
            Creation of synthetic reference using the forward solver and a scalar field  $\alpha^{\mathrm{ref}}$ manipulating the eddy viscosity $\nu_t$.
    }
    \label{fig:synthetic ref alpha ref creation}
\end{figure}

The reference divergence-free forcing is chosen as
\begin{equation}
    \label{eq:curl psi synth ref}
    \begin{split}
      & \left( \nabla \times \psi^\mathrm{ref} \right)_x (x,y) = 
    \begin{cases}
        0.01\sqrt{(x-1)^2+4y^2} -0.05, & \sqrt{(x-1)^2+4y^2} < 5 \\
        0, & \text{otherwise}
    \end{cases} \\
      &\left( \nabla \times \psi^\mathrm{ref} \right)_y (x,y) = 
    \begin{cases}
        0.0175\sqrt{x^2+16(y-1)^2} -0.07, & \sqrt{x^2+16(y-1)^2} < 4 \\
        -0.0175\sqrt{x^2+16(y+1)^2} +0.07, & \sqrt{x^2+16(y+1)^2} < 4 \\
        0, & \text{otherwise}
    \end{cases}  
    \end{split}
\end{equation}
for the $x$ component and $y$ component, respectively, as depicted in Fig.~\ref{fig:synthetic ref curl psi creation}.
The reference scalar multiplier field for the eddy viscosity is chosen as
\begin{equation}
    \label{eq:alpha nut ref}
    \alpha^\mathrm{ref} (x,y)= 
    \begin{cases}
        1-0.02\left(5-\sqrt{(x-1)^2+4y^2}\right) +0.019, & \sqrt{(x-1)^2+4y^2} < 5 \\
        1, & \text{otherwise}
    \end{cases}
\end{equation}
which is illustrated in Fig.~\ref{fig:synthetic ref alpha ref creation}.


\end{document}